\documentclass[aps, prd, reprint, letterpaper, preprintnumbers, floatfix, superscriptaddress]{revtex4-1} 
\usepackage{amsmath}
\usepackage{amssymb}
\usepackage[australian]{babel}
\usepackage{braket}
\usepackage{graphicx}
\usepackage{hyperref}
\usepackage{mathtools}
\usepackage{nicefrac}
\usepackage{siunitx}
\usepackage[caption=false]{subfig}
\usepackage[normalem]{ulem}
\usepackage{wasysym}
\usepackage{xcolor}

\usepackage{amsmath}
\usepackage{amssymb}
\usepackage{amsfonts}
\usepackage{amsthm}
\usepackage{xfrac}

\allowdisplaybreaks[1]
\newcommand{\conteqn}{\nonumber\\*}
\newcommand{\neweqn}{\\}

\newcommand{\ee}[0]{\mathrm{e}}
\newcommand{\ii}[0]{i}

\newcommand{\identity}[0]{\mathbb{I}}

\newcommand{\vect}[1]{\bm{#1}}
\newcommand{\unitvect}[1]{\smash{\widehat{\vect{#1}}}\;\!\vphantom{#1}}

\newcommand{\clovercoeff}{C_{SW}}

\newcommand{\dipoleYIntercept}{G_0}
\newcommand{\dipoleScale}{\Lambda}

\newcommand{\temporalsites}{N_t}

\newcommand{\current}{\mathcal{J}}

\newcommand{\projpm}{\projector{\pm\vect{p}}}

\newcommand{\spinhalf}[0]{spin-\sfrac{1}{2}}

\newcommand{\adjoint}[1]{\smash{\overline{#1}}\vphantom{#1}}

\newcommand{\transpose}{^\top}

\DeclareMathOperator{\Tr}{Tr}

\DeclareMathOperator{\sign}{sign}

\newcommand{\definedby}{\equiv}

\newcommand{\backderivLDr}[0]{{\overleftarrow{\nabla}^{{\rho}}}}

\newcommand{\deltaLDmLDn}[0]{{\delta^{{\mu}{\nu}}}}
\newcommand{\energySi}[1]{{E^{{\alpha}}(#1)}}
\newcommand{\epsilonCaCbCc}[0]{{\epsilon^{{a}{b}{c}}}}
\newcommand{\ffDipole}[1][Q^2]{{G_{{D}}(#1)}}
\newcommand{\ffDiracSi}[1][Q^2]{{F^{{\alpha}}_{{1}}(#1)}}
\newcommand{\ffDirac}[1][Q^2]{{F_{{1}}(#1)}}
\newcommand{\ffElectricSdblsa}[1][Q^2]{{G^{{u_{\indSprsa}}}_{{E}}(#1)}}
\newcommand{\ffElectricSi}[1][Q^2]{{G^{{\alpha}}_{{E}}(#1)}}
\newcommand{\ffElectricSnesa}[1][Q^2]{{G^{{n^*_1}}_{{E}}(#1)}}
\newcommand{\ffElectricSprsa}[1][Q^2]{{G^{{p^*_1}}_{{E}}(#1)}}
\newcommand{\ffElectricSsingsa}[1][Q^2]{{G^{{d_{\indSprsa}}}_{{E}}(#1)}}
\newcommand{\ffElectric}[1][Q^2]{{G_{{E}}(#1)}}

\newcommand{\ffMagneticSdblsa}[1][Q^2]{{G^{{u_{\indSprsa}}}_{{M}}(#1)}}
\newcommand{\ffMagneticSi}[1][Q^2]{{G^{{\alpha}}_{{M}}(#1)}}
\newcommand{\ffMagneticSsingsa}[1][Q^2]{{G^{{d_{\indSprsa}}}_{{M}}(#1)}}
\newcommand{\ffMagnetic}[1][Q^2]{{G_{{M}}(#1)}}
\newcommand{\ffPauliSi}[1][Q^2]{{F^{{\alpha}}_{{2}}(#1)}}
\newcommand{\ffPauli}[1][Q^2]{{F_{{2}}(#1)}}
\newcommand{\forwderivLDr}[0]{{\overrightarrow{\nabla}^{{\rho}}}}
\newcommand{\gammaLCk}[0]{{\gamma^{{k}}}}
\newcommand{\gammaLUfive}[0]{{\gamma^{{5}}}}
\newcommand{\gammaLUfour}[0]{{\gamma^{{4}}}}
\newcommand{\gammaLUm}[0]{{\gamma^{{\mu}}}}
\newcommand{\gevectleftSiOi}[0]{{v^{{\alpha}}_{{i}}}}
\newcommand{\gevectrightSiOj}[0]{{u^{{\alpha}}_{{j}}}}

\newcommand{\indSdblp}[0]{{u_{\indSprp}}}
\newcommand{\indSdblsa}[0]{{u_{\indSprsa}}}
\newcommand{\indSdblsb}[0]{{u_{\indSprsb}}}
\newcommand{\indSdbl}[0]{{u_{\indSpr}}}

\newcommand{\indSi}[0]{{\alpha}}

\newcommand{\indSnep}[0]{{n^{\prime}}}
\newcommand{\indSnesa}[0]{{n^*_1}}

\newcommand{\indSnp}[0]{{N^{\prime}\!}}
\newcommand{\indSnsa}[0]{{N^*_1\!}}
\newcommand{\indSnsb}[0]{{N^*_2\!}}

\newcommand{\indSprp}[0]{{p^{\prime}}}
\newcommand{\indSprsa}[0]{{p^*_1}}
\newcommand{\indSprsb}[0]{{p^*_2}}
\newcommand{\indSpr}[0]{{p}}

\newcommand{\indSsingp}[0]{{d_{\indSprp}}}
\newcommand{\indSsingsa}[0]{{d_{\indSprsa}}}
\newcommand{\indSsingsb}[0]{{d_{\indSprsb}}}
\newcommand{\indSsing}[0]{{d_{\indSpr}}}

\newcommand{\indexLCk}[1]{{{#1}^{{k}}}}
\newcommand{\indexLDe}[1]{{{#1}^{{\eta}}}}
\newcommand{\indexLDl}[1]{{{#1}^{{\lambda}}}}
\newcommand{\indexLDm}[1]{{{#1}^{{\mu}}}}
\newcommand{\indexLDn}[1]{{{#1}^{{\nu}}}}
\newcommand{\indexLDr}[1]{{{#1}^{{\rho}}}}
\newcommand{\indexLDs}[1]{{{#1}^{{\sigma}}}}
\newcommand{\interpOi}[2]{{\chi_{{i}}(#2)}}
\newcommand{\interpPEVAOip}[2]{{\chi_{{#1}\,{i'}}(#2)}}
\newcommand{\interpPEVAOi}[2]{{\chi_{{#1}\,{i}}(#2)}}
\newcommand{\interpbarPEVAOi}[2]{{\adjoint{\chi}_{{#1}\,{i}}(#2)}}

\newcommand{\interpoptPEVASi}[2]{{\phi^{{\alpha}\,{}}_{{#1}}(#2)}}
\newcommand{\interpoptbarPEVASi}[2]{{\adjoint{\phi}^{{\alpha}\,{}}_{{#1}}(#2)}}

\newcommand{\massPhysSpr}[0]{{m^{{\mathrm{{phys}}}\,{p}}}}
\newcommand{\massSi}[0]{{m^{{\alpha}}}}

\newcommand{\massSnsa}[0]{{m^{{N^*_1\!}}}}
\newcommand{\massSnsb}[0]{{m^{{N^*_2\!}}}}
\newcommand{\massSn}[0]{{m^{{N}}}}

\newcommand{\mmEff}[0]{{\mu_{{\mathrm{{eff}}}}}}
\newcommand{\mmSdblsa}[0]{{\mu^{{u_{\indSprsa}}}}}
\newcommand{\mmSnep}[0]{{\mu^{{n^{\prime}}}}}
\newcommand{\mmSnesa}[0]{{\mu^{{n^*_1}}}}
\newcommand{\mmSnesb}[0]{{\mu^{{n^*_2}}}}
\newcommand{\mmSne}[0]{{\mu^{{n}}}}
\newcommand{\mmSprp}[0]{{\mu^{{p^{\prime}}}}}
\newcommand{\mmSprsa}[0]{{\mu^{{p^*_1}}}}
\newcommand{\mmSprsb}[0]{{\mu^{{p^*_2}}}}
\newcommand{\mmSpr}[0]{{\mu^{{p}}}}
\newcommand{\mmSsingsa}[0]{{\mu^{{d_{\indSprsa}}}}}
\newcommand{\mm}[0]{{\mu}}
\newcommand{\nucleoninterpOne}[0]{{\chi_{{1}}}}
\newcommand{\nucleoninterpTwo}[0]{{\chi_{{2}}}}
\newcommand{\projectorLCk}[0]{{\Gamma^{{k}}}}
\newcommand{\projectorLUe}[0]{{\Gamma^{{\eta}}}}
\newcommand{\projectorLUfour}[0]{{\Gamma^{{4}}}}
\newcommand{\projectorLUn}[0]{{\Gamma^{{\nu}}}}
\newcommand{\projectorLUs}[0]{{\Gamma^{{\sigma}}}}
\newcommand{\projector}[2][]{{\Gamma^{{#1}}_{\!{#2}}}}

\newcommand{\quarka}[0]{{q}}

\newcommand{\quarkbara}[0]{{\adjoint{q}}}

\newcommand{\quarkdownCb}[0]{{d^{{b}}}}

\newcommand{\quarkupCa}[0]{{u^{{a}}}}
\newcommand{\quarkupCc}[0]{{u^{{c}}}}

\newcommand{\rsqElectricSnep}[0]{{\braket{r^2}^{{n^{\prime}}}_{{E}}}}
\newcommand{\rsqElectricSnesa}[0]{{\braket{r^2}^{{n^*_1}}_{{E}}}}
\newcommand{\rsqElectricSnesb}[0]{{\braket{r^2}^{{n^*_2}}_{{E}}}}
\newcommand{\rsqElectricSne}[0]{{\braket{r^2}^{{n}}_{{E}}}}
\newcommand{\rsqElectricSprp}[0]{{\braket{r^2}^{{p^{\prime}}}_{{E}}}}
\newcommand{\rsqElectricSprsa}[0]{{\braket{r^2}^{{p^*_1}}_{{E}}}}
\newcommand{\rsqElectricSprsb}[0]{{\braket{r^2}^{{p^*_2}}_{{E}}}}
\newcommand{\rsqElectricSpr}[0]{{\braket{r^2}^{{p}}_{{E}}}}
\newcommand{\rsqMagneticSdblsa}[0]{{\braket{r^2}^{{u_{\indSprsa}}}_{{M}}}}
\newcommand{\rsqMagneticSnep}[0]{{\braket{r^2}^{{n^{\prime}}}_{{M}}}}
\newcommand{\rsqMagneticSnesa}[0]{{\braket{r^2}^{{n^*_1}}_{{M}}}}
\newcommand{\rsqMagneticSnesb}[0]{{\braket{r^2}^{{n^*_2}}_{{M}}}}
\newcommand{\rsqMagneticSne}[0]{{\braket{r^2}^{{n}}_{{M}}}}
\newcommand{\rsqMagneticSprp}[0]{{\braket{r^2}^{{p^{\prime}}}_{{M}}}}
\newcommand{\rsqMagneticSprsa}[0]{{\braket{r^2}^{{p^*_1}}_{{M}}}}
\newcommand{\rsqMagneticSprsb}[0]{{\braket{r^2}^{{p^*_2}}_{{M}}}}
\newcommand{\rsqMagneticSpr}[0]{{\braket{r^2}^{{p}}_{{M}}}}
\newcommand{\rsqMagneticSsingsa}[0]{{\braket{r^2}^{{d_{\indSprsa}}}_{{M}}}}
\newcommand{\rsqMagnetic}[0]{{\braket{r^2}_{{M}}}}
\newcommand{\sigmaLUmLUn}[0]{{\sigma^{{\mu}{\nu}}}}
\newcommand{\sigmaLUrLUm}[0]{{\sigma^{{\rho}{\mu}}}}
\newcommand{\spinorSi}[1][s]{{u^{{\alpha}}{(p, #1)}}}

\newcommand{\spinorpbarSi}[1][s']{{\adjoint{u}^{{\alpha}}{(p', #1)}}}

\newcommand{\threecfSi}[6][]{{\mathcal{G}^{{3}}_{{#1}}(#2\,;#3\,,#4\,;#5\,,#6\,;\indSi{})}}

\newcommand{\threecfprojSi}[7][]{{G^{{3}}_{{#1}}(#7\,;#2\,;#3\,,#4\,;#5\,,#6\,;\indSi{})}}

\newcommand{\twocfprojPEVAOiOj}[2]{{G_{{i}{j}}(#1\,;#2)}}

\newcommand{\twocfprojPEVASi}[2]{{G(#1\,;#2\,;\indSi{})}}

\newcommand{\twocfprojPEVA}[2]{{G(#1\,;#2)}}

\newcommand{\vectorcurrentLUl}[1][]{{j^{{\lambda}}_{{#1}}}}
\newcommand{\vectorcurrentLUm}[1][]{{j^{{\mu}}_{{#1}}}}
\newcommand{\vectorcurrentLUr}[1][]{{j^{{\rho}}_{{#1}}}}

\allowdisplaybreaks[1]

\begin{document}

\preprint{
\vbox{
\hbox{ADP-19-10/T1090}
}}

\title{Elastic Form Factors of Nucleon Excitations in Lattice QCD}
\date{13 May 2020}

\author{Finn~M.~Stokes}
\affiliation{Special Research Centre for the Subatomic Structure of
  Matter,\\Department of Physics, University of Adelaide, South
  Australia 5005, Australia}
\affiliation{J\"ulich Supercomputing Centre, Institute for Advanced Simulation,\\
  Forschungszentrum J\"ulich, J\"ulich D-52425, Germany}
\author{Waseem~Kamleh}
\author{Derek~B.~Leinweber}
\affiliation{Special Research Centre for the Subatomic Structure of
  Matter,\\Department of Physics, University of Adelaide, South
  Australia 5005, Australia}

\begin{abstract}
    First principles calculations of the form factors
    of baryon excitations are now becoming accessible through
    advances in Lattice QCD techniques. In this paper, we explore
    the utility of the parity-expanded variational analysis (PEVA)
    technique in calculating the Sachs electromagnetic form factors
    for excitations of the proton and neutron. We study the two
    lowest-lying odd-parity excitations and demonstrate that at
    heavier quark masses, these states are dominated by behaviour
    consistent with constituent quark models for the \(N^*(1535)\)
    and \(N^*(1650)\), respectively. We also study the lowest-lying
    localised even-parity excitation, and find that its form factors
    are consistent with a radial excitation of the ground state
    nucleon.  A comparison of the results from the PEVA technique
    with those from a conventional variational analysis exposes the
    necessity of the PEVA approach in baryon excited-state studies.
\end{abstract}

\maketitle

\section{Introduction\label{sec:excitations:introduction}}
Investigating the structure of hadronic excited states is recognised as an important frontier in
the field of nonperturbative QCD.  At present, very little is known about how QCD composes the
structure of these excitations.  With regard to the excitations of the nucleon investigated herein,
most of our intuition is based on models of QCD as opposed to QCD itself.  The intriguing question
is, how does the quantum field theory of QCD construct these states and how does this composition
compare with the expectations of current models?  Can something as simple as a constituent quark
model capture the essence of these states?  What role do meson-baryon dressings play in describing
these states?  Our aim is to address these most fundamental questions by examining the
electromagnetic structure of nucleon excited states as observed in lattice QCD.  The results are
fascinating, validating quark model predictions in some cases and demanding a more important role
for meson-baryon interactions in others.

A longer term goal is to confront experiment.  While experimental measurements of resonance
transition amplitudes\index{form factors!transition} have been made, it is much harder to measure
elastic form factors\index{form factors} in the resonance regime.  This is because elastic form
factors parameterise interactions where both the initial and final state are the same. To measure
them for an (unstable) resonance, one needs to first produce that resonance, and then probe it
during the extremely short time window before it decays.  On the other hand, the transition form
factors parameterise the transformation of one state into another. We can probe a stable target
such as a ground state proton and measure how it is excited into the unstable resonance of interest
through an examination of its decay products.

It has been suggested that the magnetic dipole moment\index{magnetic moment}
of the \(N^*(1535)\) resonance\index{N*1535@\(N^*(1535)\)} could be measured through the
\(\gamma p \to \gamma \eta p\) process~\cite{Chiang:2002ah} using
the Crystal Barrel/TAPS detector at ELSA or Crystal Ball @ MAMI, but this
measurement has yet to be realised. The difficulty of measuring such quantities
experimentally provides the opportunity for Lattice QCD to lead experiment and
create new knowledge.

\bigskip

\subsection{Structure of Excited States}

Here we take the first step and examine the structure of nucleon
excitations as observed in the finite-volume of lattice QCD.  Using
local three-quark operators on the lattice, both the
CSSM~\cite{Mahbub:2013ala,Mahbub:2012ri} and the Hadron Spectrum
Collaboration (HSC)~\cite{Edwards:2012fx,Edwards:2011jj} observe two
low-lying odd-parity states in the resonance regimes of the
\(N^*(1535)\)\index{N*1535@\(N^*(1535)\)} and
\(N^*(1650)\)\index{N*1650@\(N^*(1650)\)}.

As these finite-volume states have good overlap with local three-quark
operators, we wish to examine the extent to which these states,
created in relativistic quantum field theory, resemble the quark-model
states postulated to describe these resonances
\cite{Chiang:2002ah,Liu:2005wg,Sharma:2013rka}.  We anticipate the
lattice-QCD states excited by such operators to be either
quark-model-like states dressed by a meson cloud, similar to the
ground-state nucleon, or perhaps bound meson-baryon molecular states,
such as the \(\Lambda(1405)\)~\cite{Hall:2014uca}.

In Ref.~\cite{Stokes:2018emx}, we presented a method for extracting the form factors\index{form
  factors} of a baryonic state on the lattice using the parity-expanded variational analysis
(PEVA)\index{PEVA} technique, and established its effectiveness for accessing the structure of the
ground-state nucleon. We now use this method (as summarised in Section~\ref{sec:peva}) to
investigate the structure of the excitations\index{excited states} of the proton and neutron
observed in the finite volume of lattice QCD\@.

In this paper we present a determination of the Sachs electric and magnetic
form factors for three \spinhalf{} nucleon eigenstates on the lattice.
Two of these states are negative-parity nucleon excitations, which we label
\(\indSnsa\) (or \(\indSprsa\) for the proton excitation and \(\indSnesa\) for
the neutron excitation), and \(\indSnsb\) (or the equivalent
labels for the excited proton and neutron). The remaining eigenstate is a
positive-parity excitation, and is denoted \(\indSnp\), \(\indSprp\),
or \(\indSnep\).

We compare the magnetic moments\index{magnetic moment} drawn from the negative-parity lattice-QCD
results to constituent quark model\index{constituent quark model} predictions for the magnetic
moments of the \(N^*(1535)\)\index{N*1535@\(N^*(1535)\)} and
\(N^*(1650)\)\index{N*1650@\(N^*(1650)\)} resonances~\cite{Chiang:2002ah, Liu:2005wg}. Such quark
model calculations can be extended to include effects from the pion cloud. We also compare our
lattice results to two such extensions~\cite{Liu:2005wg,Sharma:2013rka}. From these comparisons, we
make connections to the basis states to be considered in future Hamiltonian Effective Field Theory
(HEFT) analyses\index{HEFT}~\cite{Liu:2015ktc}.  Finally, we examine the extent to which the
electromagnetic form factors of the positive-parity excitation at $\sim1900$ MeV are consistent with
a constituent-quark-model radial excitation of the ground-state nucleon.

\subsection{Towards Baryon Resonance Structure}

In determining resonance properties from lattice QCD calculations, one requires a comprehensive
understanding of the spectrum of excited states in the finite periodic volume of the lattice.  This
spectrum includes all single, hybrid, and multi-particle contributions having the quantum numbers
of the resonance of interest.  This finite-volume spectrum composes the input into the L\"uscher
method~\cite{Luscher:1990ux} or its generalisations~\cite{Hall:2013qba,Li:2019qvh} which relate the
finite-volume energy levels to infinite volume momentum-dependent scattering amplitudes. The
application of these methods is a necessary step in connecting lattice QCD to the resonance
properties measured in experiment.

Obtaining an accurate determination of the finite-volume nucleon spectrum is challenging.  It
requires an extensive collection of baryon interpolating fields and robust correlation function
analysis techniques.  Many collaborations have explored the nucleon spectrum excited by local
single-particle operators~\cite{Mahbub:2010rm,Edwards:2011jj,Edwards:2012fx,Mahbub:2013ala,
  Alexandrou:2014mka,Kiratidis:2015vpa,Kiratidis:2016hda,Liu:2016uzk,Lang:2016hnn}. Hybrid nucleon
interpolators have been investigated in Ref.~\cite{Dudek:2012ag} where additional states were found
in the spectrum. Non-local multi-particle interpolating fields are necessary to quantify avoided
level crossings and determine the lattice energy eigenstates to the level of accuracy
\cite{Wilson:2015dqa} required for the implementation of the L\"uscher formalism, bringing
lattice-QCD results to experiment.  In light of these challenges, the main approach has been to
bring experimental measurements to the finite volume of the lattice
\cite{Liu:2016uzk,Liu:2016wxq,Wu:2017qve}.  It is only recently that the first applications of the
L\"uscher formalism to the lattice-baryon spectrum have
emerged~\cite{Andersen:2017una,Andersen:2019ktw}.

The computational challenges in the baryon sector contrast the tremendous progress made within the
meson sector.  For example, using the formalism for connecting precision finite-volume lattice-QCD
matrix-element calculations to the transition amplitudes of experiment
\cite{Briceno:2014uqa,Briceno:2015csa}, the resonant $\pi^+\gamma\to\pi^+\pi^0$ amplitude was first
explored in Ref.~\cite{Briceno:2015dca}.  More recently, the $\pi\pi\to\pi\gamma^\star$ amplitude
\cite{Briceno:2016kkp}, the resonant $\rho\to\pi\gamma^\star$ transition \cite{Briceno:2016kkp},
the $\pi\gamma \to \pi\pi$ transition \cite{Alexandrou:2018jbt} and $\rho$-meson radiative decay
\cite{Alexandrou:2018jbt} have been studied.  In these calculations, the finite-volume lattice matrix
elements are related to the physical momentum-transfer and energy-dependent scattering observables
of experiment.  Here a resonance appears as an enhancement as a function of the scattering energy.

A formalism for connecting the finite-volume matrix elements under investigation herein to
experiment has been presented in Refs.~\cite{Briceno:2015tza} and \cite{Baroni:2018iau}.
While our lattice-QCD formalism for matrix-element determination respects the
subtleties of the finite volume, the calculations are not sufficiently precise to engage in the
connection to experimental scattering observables.  There, one needs the contributions of
multi-particle scattering states to ensure eigenstate-projected correlation functions contain no
contaminations, to quantify the exact eigenstate energies, and to include their contributions to
resonances, which can be spread over several finite-volume energy eigenstates, particularly for
large lattice volumes where the quantised momentum spacing becomes narrow.

The calculation of lattice matrix elements for momentum-projected meson-baryon interpolators
have yet to be reported in the literature.  However, their calculation for non-forward momentum
transfers will be founded on the formalism presented and utilised herein.

\subsection{Overview}

In this paper we calculate the Sachs electric and magnetic form factors for three \spinhalf{}
nucleon excitations observed on the finite-volume lattice.
We commence with a brief summary of the parity-expanded variational analysis for matrix elements in
Sec.~\ref{sec:peva}. There, the highlights of how the PEVA projectors alter the standard formalism
is presented.
Lattice QCD gauge fields, parameters and associated analysis techniques are summarised in
Sec.~\ref{sec:techniques}.

Section~\ref{sec:excitations:neg} presents calculations of the electromagnetic form
factors for the two low-lying negative parity excitations observed on the lattice.  There, the
focus is on the utility of the PEVA formalism in removing opposite-parity
contaminations from the lattice correlation functions.  The importance of the formalism is quantified by
comparing with a conventional variational analysis where opposite-parity contaminations are not
addressed through an expansion of the correlation matrix.
Of particular note is a comparison of the magnetic moments  drawn from the negative-parity lattice-QCD
results to constituent-quark-model predictions in Sec.~\ref{sec:excitations:model}.

Finally, the electromagnetic structure of the first positive-parity excitation observed at
$\sim1900$ MeV in our lattice QCD calculations is presented in Sec.~\ref{sec:excitations:1stpos}.
The extent to which this excitation is consistent with a constituent-quark-model radial excitation
of the ground-state nucleon is of particular interest.  In accord with other studies
\cite{Roberts:2013ipa,Roberts:2013oea}, we find the structure to be consistent with a radial
excitation, further strengthening the case that the Roper resonance is not associated with a
quark-model like state \cite{Leinweber:2015kyz,Liu:2016uzk,Wu:2017qve}.

A summary and outline of future work is provided in the conclusions of
Sec.~\ref{sec:excitations:conclusion}.

\bigskip

\section{Parity Expanded Variational Analysis\label{sec:peva}}
The process of extracting finite-momentum matrix elements of baryonic excited
states via the PEVA technique is presented in full in Ref~\cite{Stokes:2018emx}. We provide
here a brief summary of this process to introduce the notation and concepts
necessary to present our results.

The idea of using operator overlaps to project onto excited states
\cite{Dudek:2009kk,Owen:2015fra,Shultz:2015pfa} and separate opposite parities
\cite{Thomas:2011rh,Padmanath:2018tuc} has also been considered in the meson sector.  In this case,
matrix elements can be extracted using standard techniques
\cite{Dudek:2009kk,Owen:2015fra,Shultz:2015pfa} as the intricacies of parity mixing within the
spinor components of the correlator are absent.  In the baryon sector, one must take the PEVA
projectors into account in identifying the appropriate Dirac-index combinations required to isolate
the covariant vertex functions and associated Sachs form factors.

We begin with a basis of \(n\) conventional spin-\nicefrac{1}{2} operators
\(\left\{\interpOi{}{x}\right\}\) that couple to the states of interest.
Adopting the Pauli representation of the gamma matrices, we introduce the PEVA
projector~\cite{Menadue:2013kfi}
\(\projpm \definedby \frac{1}{4} \left(\identity + {\gammaLUfour}\right)
\left(\identity \pm \ii {\gammaLUfive} {\gammaLCk} \indexLCk{\unitvect{p}}\right)\),
and construct a set of basis operators
\begin{subequations}
\begin{align}
    \interpPEVAOi{\pm\vect{p}}{x} &\definedby \projpm \, \interpOi{}{x}\,,\neweqn
    \interpPEVAOip{\pm\vect{p}}{x} &\definedby \pm \projpm \, \gammaLUfive{} \, \interpOi{}{x}\,.
\end{align}
\end{subequations}

We note that we use a Euclidean metric \(\deltaLDmLDn{}\), and hence there is
no need to distinguish between contravariant and covariant indices.

We then seek an optimised set of operators \(\interpoptPEVASi{\pm\vect{p}}{x}\)
that each couple strongly to a single energy eigenstate \(\indSi\). These
optimised operators are constructed as linear combinations of the basis
operators. The optimum linear combinations are found by solving a generalised
eigenvalue problem with
\(\twocfprojPEVA{\vect{p}}{t+\Delta{}t}\) and \(\twocfprojPEVA{\vect{p}}{t}\),
where the correlation matrix
\begin{align}
    &\twocfprojPEVAOiOj{\vect{p}}{t} \conteqn
    &\qquad\definedby
      \Tr\left(\sum_{\vect{x}} \ee^{-\ii \vect{p}\cdot\vect{x}}
      \braket{\Omega|\,\interpPEVAOi{\pm\vect{p}}{x}\,
        \interpbarPEVAOi{\pm\vect{p}}{0}\,|\Omega} \right)\,,
\end{align}
with \(i\) and \(j\) ranging over both the primed and unprimed operators. This
process is described in detail in Ref.~\cite{Menadue:2013kfi}.

Using the optimised operators, we can construct the eigenstate-projected
two-point correlation function
\begin{align}
    &\twocfprojPEVASi{\vect{p}}{t} \conteqn
    &\qquad\definedby
      \Tr\left(\sum_{\vect{x}} \ee^{-\ii \vect{p}\cdot\vect{x}}
      \braket{\Omega|\,\interpoptPEVASi{\pm\vect{p}}{x}\,
        \interpoptbarPEVASi{\pm\vect{p}}{0}\,|\Omega} \right) \conteqn
    &\qquad=\gevectleftSiOi(\vect{p})\,
      \twocfprojPEVAOiOj{\vect{p}}{t}\, \gevectrightSiOj(\vect{p})\,,
\end{align}
and the three point correlation functions
\begin{align}
    &\threecfSi[\pm]{\current\!}{\vect{p}'\!}{\vect{p}}{t_2}{t_1}\conteqn
    &\qquad\definedby \sum_{\vect{x}_2,\vect{x}_2} \ee^{-\ii \vect{p}'\cdot\vect{x}_2} \,
        \ee^{\ii (\vect{p}' - \vect{p})\cdot\vect{x}_1}\conteqn
    &\qquad\qquad\quad\times\braket{\Omega|\,\interpoptPEVASi{\pm\vect{p}'}{x_2}\,
        \current(x_1)\,\interpoptbarPEVASi{+\vect{p}}{0}\,|\Omega}\,,
\end{align}
where \(\current(x)\) is some current operator\index{current operator}, which is
inserted with a momentum transfer \(\vect{q} = \vect{p}' - \vect{p}\). The
consideration of \(\threecfSi[-]{\current\!}{\vect{p}'\!}{\vect{p}}{t_2}{t_1}\)
(where the sink operator uses the opposite PEVA projector sign convention to
the source operator) is required to optimise the extraction of the form factors
for general kinematics. We note that it is sufficient to consider this change
of projector for the sink operator alone, leaving the source operator as
\(\interpoptbarPEVASi{+\vect{p}}{0}\) in all cases considered.

In this paper, we investigate the electromagnetic properties of the proton and
neutron by choosing the current operator \(\current(x)\) to be the vector
current\index{vector current}. In particular, we use the \(O(a)\)-improved~\cite{Martinelli:1990ny}
conserved vector current\index{vector current}
used in Ref.~\cite{Boinepalli:2006xd},
\begin{equation}
    \vectorcurrentLUm[CI]{}(x) \definedby \vectorcurrentLUm[C]{}(x)
    + \frac{r}{2}\,a\, \quarkbara{}(x)\left(\backderivLDr{} + \forwderivLDr{}\right)
            \sigmaLUrLUm{}\,\quarka{}(x)\,,
\end{equation}
where \(r\) is the Wilson parameter, and \(\vectorcurrentLUm[C]{}(x)\) is the
standard conserved vector current for the Wilson action.

This choice of current operator gives the matrix element
\begin{align}
    &\braket{\indSi\,; p'\,; s' | \,\vectorcurrentLUm[CI]{}(0)\, | \indSi\,; p\,; s}\conteqn
    &\qquad = \sqrt{\frac{\massSi}{\energySi{\vect{p}}}} \,
              \sqrt{\frac{\massSi}{\energySi{\vect{p}'}}} \;
              \spinorpbarSi{} \conteqn
    &\qquad\qquad\quad\times \left(\gammaLUm{} \, \ffDiracSi
        - \frac{\sigmaLUmLUn{}\,\indexLDn{q}}{2 \massSi{}} \, \ffPauliSi\right)\conteqn
    &\qquad\qquad\quad\times\spinorSi{}\,,\label{eqn:formfactors:matrixelement}
\end{align}
where \(Q^2 = \vect{q}^2 - {\left(\energySi{\vect{p}'} - \energySi{\vect{p}}\right)}^2\)
is the squared four-momentum with the conventional sign, and the invariant scalar
functions \(\ffDirac\) and \(\ffPauli\) are respectively the
Dirac\index{form factors!Dirac} and Pauli\index{form factors!Pauli} form factors.
Here \(\spinorSi{}\) is the spinor for the lattice eigenstate \(\indSi{}\) moving
with momentum \(p\) and spin \(s\). As \(\indSi{}\) is an eigenstate of the
lattice Hamiltonian, this spinor takes the form of a conventional single
particle spinor with the centre-of-momentum energy,
\(E_{CM} = \sqrt{\energySi{\vect{p}}^2-\vec{p}^2}\) playing the role of mass
in the finite volume.
The states considered in this work display an energy-momentum relation consistent
with a single particle dispersion relation~\cite{Menadue:2013kfi}.
As such, \(E_{CM} \approx \massSi{}\) throughout this work.
However, the techniques
presented respect the subtleties of the finite volume and are applicable for states where \(E_{CM} \ne \massSi{}\).

To extract our desired signal from this spinor structure, we can take the spinor trace
with some spin-structure projector\index{spin-structure projector}
\(\projector{S}\). This trace is then called the
spinor-projected three-point correlation function\index{correlation function!three point}
\begin{align}
    &\threecfprojSi[\pm]{\vectorcurrentLUm[CI]{}}{\vect{p}'\!}{\vect{p}}{t_2}{t_1}{\projector{S}}\conteqn
        &\qquad\definedby \Tr\!\left(\,\projector{S}\,
        \threecfSi[\pm]{\vectorcurrentLUm[CI]{}}{\vect{p}'\!}{\vect{p}}{t_2}{t_1}\, \right)\,.
\end{align}

These spinor-projected correlation functions\index{correlation function!three point} have a
nontrivial time dependence, which can be removed by constructing the ratio~\cite{Leinweber:1990dv}\index{ratio}
\begin{widetext}
\begin{align}\label{eqn:formfactors:ratio}
    R_{\pm}(\vect{p}', \vect{p}\,; \alpha\,; r, s) &\definedby \, \sqrt{\left|\frac{
        \indexLDm{r} \, \threecfprojSi[\pm]{\vectorcurrentLUm[CI]{}}{\vect{p}'\!}{\vect{p}}{t_2}{t_1}{\indexLDn{s} \, \projectorLUn{}} \,
        \indexLDr{r} \, \threecfprojSi[\pm]{\vectorcurrentLUr[CI]{}}{\vect{p}}{\vect{p}'\!}{t_2}{t_1}{\indexLDs{s} \, \projectorLUs{}}}{
        \twocfprojPEVASi{\vect{p}'}{t_2} \, \twocfprojPEVASi{\vect{p}}{t_2}}\right|} \conteqn
    &\quad \times \sign\!\left(
         \indexLDl{r} \, \threecfprojSi[\pm]{\vectorcurrentLUl[CI]{}}{\vect{p}'\!}{\vect{p}}{t_2}{t_1}{\indexLDe{s} \, \projectorLUe{}}\right) \,,
\end{align}\index{R@\(R_{\pm}(\vect{p}', \vect{p}\,; \alpha\,; r, s)\)|see {ratio}}\end{widetext}
where \(\projectorLUfour{} = (\identity + \gammaLUfour{}) / 2\) and
\(\projectorLCk{} = (\identity + \gammaLUfour{}) (\ii \, \gammaLUfive{} \, \gammaLCk{}) / 2\) form
the basis for the spin projectors we use, and \(\indexLDm{r}\) and \(\indexLDm{s}\) are
coefficients selected to determine the form factors.

We can then define the reduced ratio\index{reduced ratio},
\begin{align}
    &\adjoint{R}_{\pm}(\vect{p}', \vect{p}\,; \alpha\,; r, s) \conteqn
    &\qquad\definedby
        \sqrt{\frac{2 \energySi{\vect{p}}}{\energySi{\vect{p}}+\massSi{}}} \,
        \sqrt{\frac{2 \energySi{\vect{p}'}}{\energySi{\vect{p}'}+\massSi{}}} \conteqn
    &\qquad\qquad\quad\times R_{\pm}(\vect{p}', \vect{p}\,; \alpha\,; r, s) \,.
\end{align}\index{Rbar@\(\adjoint{R}_{\pm}(\vect{p}', \vect{p}\,; \alpha\,; r, s)\)|see {reduced ratio}}

By investigating the \(\indexLDm{r}\) and \(\indexLDs{s}\) dependence of this
ratio, we find that the clearest signals are given by
\begin{subequations}\label{eqn:formfactors:selectedratios}
\begin{align}
    R^{T}_{\pm} &= \frac{2}{1 \pm\, \unitvect{p} \cdot \unitvect{p}'} \;
        \adjoint{R}_{\pm}\left(\vect{p}', \vect{p}\,; \alpha\,; (1, \vect{0}), (1, \vect{0})\right)
        \,, \neweqn
    R^{S}_{\mp} &= \frac{2}{1 \pm\, \unitvect{p} \cdot \unitvect{p}'} \;
        \adjoint{R}_{\mp}\left(\vect{p}', \vect{p}\,; \alpha\,; (0, \unitvect{r}),(0, \unitvect{s})\right)\,,
\end{align}
\end{subequations}
where \(\unitvect{s}\) is chosen such that
\(\vect{p} \cdot \unitvect{s} = 0 = \vect{p}' \cdot \unitvect{s}\), \(\unitvect{r}\) is
equal to \(\unitvect{q} \times \unitvect{s}\), and the sign \(\pm\) in
Eq.~\eqref{eqn:formfactors:selectedratios} is chosen such that
\(1 \pm \unitvect{p} \cdot \unitvect{p}'\) is maximised. This choice maximises
the signal in the lattice determination of the correlation function ratios.

We can then find the Sachs electric and magnetic form factors\index{form factors!Sachs},
\begin{subequations}
\begin{align}
    \ffElectricSi{} &\definedby \ffDiracSi{} - \frac{Q^2}{{\left(2\massSi{}\right)}^2} \, \ffPauliSi{}\,, \neweqn
    \ffMagneticSi{} &\definedby \ffDiracSi{} + \ffPauliSi{}\,,
\end{align}
\end{subequations}
through appropriate linear combinations of \(R^{T}_{\pm}\) and \(R^{S}_{\mp}\).

We have shown how the PEVA technique can be applied to the calculation of
elastic baryon form
factors for arbitrary kinematics. We now proceed to investigate the Sachs
electric and magnetic form factors of several excitations of the nucleon.

\section{Lattice QCD Parameters and Analysis Techniques\label{sec:techniques}}
\begin{table*}[hbt]
    \caption{\label{tab:formfactors:ensembles}Details of the gauge field ensembles
        used in this analysis.
        For each ensemble we list both the pion mass given in Ref.~\cite{Aoki:2008sm},
        with the lattice spacing set by hadronic inputs, and our determination
        of the the squared pion mass with the lattice spacing listed in the
        table, which is set by the Sommer parameter with
        \(r_0 = \SI{0.4921 \pm 0.0064}{\femto\meter}\)~\cite{Aoki:2008sm}.         }
    \sisetup{%
    table-number-alignment = center,
    table-figures-integer = 1,
    table-figures-decimal = 0,
    table-figures-uncertainty = 0,
}
\begin{tabular*}{\linewidth}{%
@{\extracolsep{\fill}}
        S[table-figures-integer = 3, table-figures-decimal = 0, table-figures-uncertainty = 0]
        S[table-figures-integer = 1, table-figures-decimal = 4, table-figures-uncertainty = 2]
        S[table-figures-integer = 1, table-figures-decimal = 4, table-figures-uncertainty = 3]
        S[table-figures-integer = 3, table-figures-decimal = 0, table-figures-uncertainty = 0]
        S[table-figures-integer = 1, table-figures-decimal = 0, table-figures-uncertainty = 0]
}
    \toprule{} \vspace{-9pt}\\
    {PACS-CS \(m_{\pi} \, / \,\si{\mega\electronvolt}\)} & {\(a \, / \, \si{\femto\meter}\)} & {\(m_{\pi}^2 \, / \, \si{\giga\electronvolt^2}\)} & {\# conf.} & {\# src per conf.} \\
    \colrule{} \vspace{-9pt}\\
    702 & 0.1022 \pm 0.0015 & 0.3884 \pm 0.0113 & 399 & 1 \\
    570 & 0.1009 \pm 0.0015 & 0.2654 \pm 0.0081 & 397 & 1 \\
    411 & 0.0961 \pm 0.0013 & 0.1525 \pm 0.0043 & 449 & 2 \\
    296 & 0.0951 \pm 0.0013 & 0.0784 \pm 0.0025 & 400 & 2 \\
    156 & 0.0933 \pm 0.0013 & 0.0285 \pm 0.0012 & 197 & 4 \\
    \botrule{}
\end{tabular*}
\end{table*}

\subsection{Gauge Field Configurations}

The results presented in this paper are  calculated on the PACS-CS
\((2+1)\)-flavour full-QCD ensembles~\cite{Aoki:2008sm}, made available through the
ILDG~\cite{Beckett:2009cb}. These ensembles use a \(32^3 \times 64\) lattice, and
employ a renormalisation-group improved Iwasaki gauge action\index{action:improved} with
\(\beta = 1.90\) and non-perturbatively \(O(a)\)-improved Wilson quarks, with
\(\clovercoeff = 1.715\). We use five ensembles, with stated pion masses
from \(m_{\pi}=\SI{702}{\mega\electronvolt}\) to \(\SI{156}{\mega\electronvolt}\)~\cite{Aoki:2008sm},
and set the scale using the Sommer parameter with \(r_0 = \SI{0.4921 \pm 0.0064}{\femto\meter}\)~\cite{Aoki:2008sm}. More details of the individual ensembles are presented in Table~\ref{tab:formfactors:ensembles},
including the squared pion masses in the Sommer scale. When fitting
correlators, the \(\chi^2 / \mathrm{dof}\) is calculated with the full covariance
matrix, and the \(\chi^2\) values of all fits are consistent with an appropriate
\(\chi^2\) distribution.

The three heaviest pion masses available among these ensembles span
\(m_\pi = \SI{411}{\mega\electronvolt}\)--\(\SI{702}{\mega\electronvolt}\),
a typical range for contemporary studies of baryon excitations. As such, these masses are
appropriate for this world-first study of the electromagnetic structure of
nucleon excitations in lattice QCD.
In presenting our discoveries, we will focus on the results
at these three heaviest pion masses.
There are two lighter masses, at
\(m_\pi = \SI{156}{\mega\electronvolt}\) and \(\SI{296}{\mega\electronvolt}\).
These approach the physical point, presenting a significant challenge in terms of
gauge noise and computational cost, but offer the possibility of insight into
important chiral physics.

\subsection{Conventional and PEVA Techniques\label{subsec:conv:def}}

For the variational analyses in this paper, we begin with the same
eight-interpolator basis as in Ref.~\cite{Stokes:2018emx}, in which we studied
the electromagnetic form factors of the ground-state nucleon. This basis is
formed from the conventional \spinhalf{} nucleon interpolators
\begin{align}
    \nucleoninterpOne = &\epsilonCaCbCc [{\quarkupCa}\transpose \, (C\gammaLUfive) \, \quarkdownCb] \quarkupCc\,,\ \text{and}\conteqn
    \nucleoninterpTwo = &\epsilonCaCbCc [{\quarkupCa}\transpose \, (C) \, \quarkdownCb] \gammaLUfive \quarkupCc\,,
\end{align}
with \num{16}, \num{35}, \num{100}, or \num{200} sweeps of gauge-invariant Gaussian
smearing\index{gauge-invariant Gaussian smearing}~\cite{Gusken:1989qx} with a smearing fraction of
\(\alpha = 0.7\), applied at the quark source and sinks in creating the propagators.  Before
performing the Gaussian smearing, the gauge links to be used are smoothed by applying four sweeps
of three-dimensional isotropic stout-link smearing~\cite{Morningstar:2003gk} with \(\rho=0.1\).
We will refer to analyses based on this $8\times 8$ correlation matrix, without opposite-parity
interpolators, as the {\sl conventional variational analysis}.
For the {\sl PEVA technique}, this basis is
expanded to sixteen operators as described in Section~\ref{sec:peva}.

We study the first three excitations extracted by this basis, consisting of one positive-parity
state and two negative-parity states.  As we will see in the results presented below, the
PEVA\index{PEVA} technique is very important in correctly extracting form factors of these
excitations.\\

\subsection{Three-point Function Techniques}

To extract the form factors, we fix the source at time slice \(\temporalsites/4=\num{16}\)
relative to a fixed boundary condition in time, and (utilising
the sequential source technique~\cite{Bernard:1985ss}) invert through the current,
fixing the current insertion at time slice \num{21}. We choose time slice \num{21}
by inspecting the two point correlation functions associated with each state and
observing that excited-state contaminations in the eigenstate-projected
correlators are suppressed by time slice
\num{21}. This is evaluated by fitting the effective mass\index{effective energy} in this region to a
single state ansatz verifying that the full covariant \(\chi^2 / \text{dof}\)
is satisfactory. We then extract the form
factors as outlined in Section~\ref{sec:peva} for every
possible sink time and once again look for a plateau consistent with a
single-state ansatz.

\subsection{Multi-Particle Scattering State Contributions \label{subsec:multi-part}}

The quasi-local operators used to excite the states of interest do not have good overlap with
multi-particle scattering eigenstates.  As such, a particular concern in this analysis is the
possibility of contamination of our correlation functions by nearby multi-particle scattering
states that have not been isolated in the current correlation-matrix analysis using local
operators.

Fortunately, significant mixing of one- and two-particle basis states gives rise to
avoided level crossings creating a large energy separation between the lattice energy
eigenstates. This difference in energies leads to a rejection of single-state ansatz fits
\cite{Kiratidis:2015vpa} signified by a large covariance-matrix \(\chi^2 / \mathrm{dof}\).

However, when the mixing of the basis states is small, the avoided level crossing effects are
significantly reduced. This can allow lattice eigenstates with closely spaced energy levels. Such
nearby eigenstates are more problematic as simple Euclidean time evolution cannot expose separate
states.  If one is interested only in the energies of the eigenstates, the mixing can shift the
observed energies by an small amount, typically within the width of the associated resonance.  On
the other hand, it is these subtle shifts that are central to the L\"uscher formalism
\cite{Wilson:2015dqa}.
This issue of subtle state mixing applies to form factors in a more significant manner as the form
factors of the scattering states may differ significantly from the energy eigenstates having good
overlap with the local interpolators.  Thus it is important to estimate the extent of this mixing.

\begin{table}[bt]
    \caption{\label{tab:two-particle-threshold} Sommer-scale masses and two-particle
      infinite-volume threshold energies in units of \(\si{\giga\electronvolt}\) are compared with
      the masses of the first, $N_1^*$, and second, $N_2^*$, negative-parity states observed on the
      lattice.
}
    \sisetup{%
    table-number-alignment = center,
    table-figures-integer = 1,
    table-figures-decimal = 0,
    table-figures-uncertainty = 0,
}
\begin{tabular*}{\linewidth}{%
@{\extracolsep{\fill}}
        S[table-figures-integer = 1, table-figures-decimal = 3, table-figures-uncertainty = 1]
        S[table-figures-integer = 1, table-figures-decimal = 2, table-figures-uncertainty = 1]
        S[table-figures-integer = 1, table-figures-decimal = 2, table-figures-uncertainty = 1]
        S[table-figures-integer = 1, table-figures-decimal = 2, table-figures-uncertainty = 2]
        S[table-figures-integer = 1, table-figures-decimal = 2, table-figures-uncertainty = 1]
}
    \toprule{} \vspace{-9pt}\\
    {\(m_{\pi} / \si{\giga\electronvolt}\)} & {\(\massSn{} \! / \si{\giga\electronvolt}\)} & {\(\left(\massSn{} \! + m_{\pi} \right) \! / \si{\giga\electronvolt}\)} & {\(\massSnsa{} \! / \si{\giga\electronvolt}\)} & {\(\massSnsb{} \! / \si{\giga\electronvolt}\)} \\
    \colrule{} \vspace{-9pt}\\
    0.623 \pm 0.009 & 1.41 \pm 0.01 & 2.03 \pm 0.01 & 1.90 \pm 0.04 & 1.95 \pm 0.02 \\
    0.515 \pm 0.008 & 1.27 \pm 0.01 & 1.78 \pm 0.01 & 1.78 \pm 0.05 & 1.82 \pm 0.02 \\
    0.391 \pm 0.006 & 1.15 \pm 0.01 & 1.54 \pm 0.01 & 1.71 \pm 0.03 & 1.77 \pm 0.03 \\
    0.280 \pm 0.004 & 1.06 \pm 0.01 & 1.34 \pm 0.01 & 1.56 \pm 0.05 & 1.75 \pm 0.09 \\
    0.169 \pm 0.004 & 1.01 \pm 0.04 & 1.18 \pm 0.04 & 1.49 \pm 0.12 & 1.55 \pm 0.08 \\
    \botrule{}
\end{tabular*}
\end{table}

For the two low-lying negative-parity states, an examination of the two-particle threshold energies
relative to the energies of the observed excitations can provide some insight.  Table
\ref{tab:two-particle-threshold} presents Sommer scale masses and two-particle infinite-volume
threshold energies which can be compared with the masses of the first and second negative-parity
states observed on the lattice \cite{Liu:2015ktc}.  While one does not have any insight into how
the $\pi N$ threshold energy gets dressed in the finite volume relative to $N_1^*$ and $N_2^*$
negative-parity states, one can see that at the largest quark mass considered, the threshold is
above the two energies observed on the lattice.  In light of the volume suppression of two-particle
couplings to local interpolating fields, scattering state contamination is not a significant
concern at this heaviest quark mass.

At the second heaviest mass considered, the situation is more complicated and one must turn to
calculations that do account for the mixing of the $\pi N$ threshold basis state with other states
in the system.  Hamiltonian effective field theory (HEFT) calculations for odd-parity nucleon
excitations can provide considerable insight \cite{Liu:2015ktc,Liu:2017wsg}.
Figures 2 and 3 of Ref.~\cite{Liu:2017wsg} indicate the $\pi N$ basis state is strongly mixed in
creating the two lowest-lying odd-parity states observed on the lattice.  Once again, we observe
that there is no low-lying $\pi N$ scattering-state contaminant.

At the middle quark mass considered, the $\pi N$ scattering state lies well below the two states
seen in lattice QCD.  Here, HEFT can provide some insight if one uses the overlap of the local bare
basis state with the energy eigenstates in HEFT as a proxy for the overlap of the local lattice
interpolating fields with the lattice energy eigenstates.
The idea is that the bare basis state in HEFT is the only localised basis state in the theory.
Drawing on chiral perturbation theory, one can show the overlap of a smeared interpolating field
with non-local momentum-projected $\pi N$ basis states is suppressed by factor of $\sim 10^{-3}$
relative to the ground state \cite{Bar:2017kxh} for our current lattice parameters.  Thus, the
optimised smeared interpolating field of lattice QCD is associated with the bare basis state of
HEFT, $\ket{ m_0 }$,
and the element $|\langle\, m_0 \,|\, E_\alpha \,\rangle|^2$ of the HEFT eigenvector governs the
relative probability of exciting eigenstate $|\, E_\alpha \,\rangle$.
With this approximation, Fig.~3 of Ref.~\cite{Liu:2017wsg} suggests a 5\% scattering-state
contribution to our projected correlators.

At the second lightest of the quark masses considered herein, the mixing of the low-lying
two-particle $\pi N$ scattering state is large enough to be quantified \cite{Mahbub:2013bba}.
While no scattering-state contamination is manifest in the eigenstate-projected correlator
associated with the \(N^*(1535)\), the correlator associated with the second negative-parity
excitation did reveal a small contamination.  By extending the Euclidean-time fit regime into the
tail of the projected two-point correlator, the high-precision analysis of
Ref.~\cite{Mahbub:2013bba} resolves a second low-lying state consistent with a $\pi N$ scattering
state at the 10\% level.  This contribution is consistent with expectations from HEFT~\cite{Liu:2017wsg}.
Our present calculation avoids the tail of this projected correlator and
our use of the single state ansatz \cite{Kiratidis:2015vpa} ensures that these contributions
to the effective energy are contained within the statistical uncertainties of the results.
However, as shown in Ref.~\cite{Stokes:2018emx}, contaminants that do not significantly
perturb the extracted mass can still have a significant effect on the extracted
form factors. As such, we must be cautious when interpreting results from this
state at this mass.

Finally, at the lightest quark mass considered, we anticipate a similar contribution from $\pi N$
scattering states.  However, statistical uncertainties at the lightest quark mass are large and we
have been unable to resolve any evidence of scattering state contamination.  As one moves towards a
precise examination of these states, one must also accommodate $K \Lambda$ and $K \Sigma$
scattering states in the analysis as the energies of these scattering-state thresholds are in the
regime of the $N^*$ states under examination at this near-physical quark mass.

With regard to the positive parity excitation examined in Sec. V, the formidable
challenge of extracting full knowledge of the many possible scattering state contributions to
the spectrum of eigenstates on the lattice is well beyond the current capabilities of the lattice
community and may only be realised with the benefit of significant algorithmic
and/or computational advances.  For
example, at the lightest quark mass considered, there is a multi-particle scattering state
associated with a nucleon plus five pions which lies below the first excitation observed on the
lattice at \(\simeq \SI{1.9}{\giga\electronvolt}\).  Again, using the overlap of the bare basis state with the energy
eigenstates in HEFT as a proxy for the overlap of the local lattice interpolators with the lattice
energy eigenstates, Figs.~3 and 5 of Ref.~\cite{Wu:2017qve} indicate the only states having
significant overlap with local interpolating fields are the states under examination herein.

This expectation is in accord with the results of Ref.~\cite{Lang:2016hnn} based on the same
PACS-CS lattices examined herein.  There Fig.~4 illustrates how the inclusion of low-lying
momentum-projected two-particle $\pi N$ and $\sigma N$ interpolators has a marginal effect on the
mass of the state determined with local interpolators alone.

In summary, the contamination of our correlation functions by nearby multi-particle scattering
states that have not been isolated in the current correlation-matrix analysis using local operators
is expected to be small.  At the heaviest quark masses, there is no issue with low-lying scattering
states.  At the lightest quark masses, a small contamination of approximately 10\% may be found in
the projected correlator of the second negative-parity excitation most associated with the
\(N^*(1650)\) resonance.  As our main focus is on the heaviest three masses and quark-model
comparisons, scattering-state contributions do not pose a significant issue in this first
examination of excited-state electromagnetic structure.  However, future calculations seeking a
quantitative connection to the scattering observables of experiment will require the inclusion of
non-local multi-particle interpolating fields.

\bigskip

\section{Negative parity excitations\label{sec:excitations:neg}}
\subsection{\texorpdfstring{\(G_E\)}{GE} for the first negative-parity excitation\label{sec:excitations:1stneg:GE}}

\begin{figure}[tbp]
    {\centering
        \includegraphics{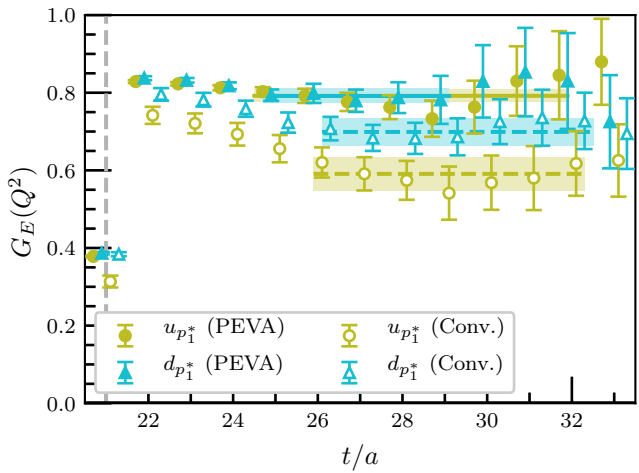}}
    \caption{\label{fig:1stneg:GE:k1p000pp100}Quark-flavour contributions to the electric form
      factor for the first negative-parity excitation of the nucleon at \(m_{\pi} =
      \SI{702}{\mega\electronvolt}\) for the lowest-momentum kinematics, providing \(Q^2 =
      \SI{0.1424 \pm 0.0041}{\giga\electronvolt^2}\).  We plot the conventional analysis with open
      markers and the new PEVA technique with filled markers. Our fits to the plateaus are
      illustrated by shaded bands, with dashed fit lines for the conventional analysis, and solid
      fit lines for PEVA\@.  The source is at time slice 16, and the current is inserted at time
      slice 21, as indicated by the vertical dashed line.  Both PEVA fits are from time slice
      \num{25}, whereas the conventional fits both start at \num{26}, and have significantly lower
      values than the PEVA fits.}
\end{figure}

\subsubsection{Quark-flavour contributions}

Beginning with the lowest-lying negative-parity excitation observed in this study, we examine how
extractions of \(\ffElectric{}\) by both the PEVA technique and the conventional analysis defined in
Sec.~\ref{subsec:conv:def} depend on the Euclidean time of the sink. In
Fig.~\ref{fig:1stneg:GE:k1p000pp100}, we plot the connected contributions to \(\ffElectric{}\) from
single quarks of unit charge for both quark flavours present in the nucleon interpolator. This plot
is at the heaviest quark mass considered with \(m_{\pi} = \SI{702}{\mega\electronvolt}\) and the
lowest-momentum kinematics of \(\vect{p} = (0,0,0)\) and \(\vect{p}' = (1,0,0)\).

We see that the
conventional extraction sits well below the PEVA extraction for all time slices
between the current insertion and the point at which the signal is lost to noise. The
conventional extraction also has a more significant time dependence than the
PEVA extraction, forcing the conventional fit one time slice later. Both of
these effects indicate that the conventional analysis is affected by
opposite-parity contaminations, which are having a significant effect
on the extracted form factor, introducing a systematic error of
\SI{12 \pm 4}{\percent}
for the singly represented quark flavour and
\SI{25 \pm 5}{\percent}
for the doubly represented flavour.

The lighter pion masses show a similar behaviour. The conventional
analysis consistently has a plateau which starts later than the PEVA approach
and sits significantly lower. For example, at \(m_{\pi} = \SI{411}{\mega\electronvolt}\),
the magnitudes of the conventional plateaus with these low-momentum kinematics
are systematically underestimated by
\SI{19 \pm 6}{\percent}
and \SI{26 \pm 6}{\percent}
for the singly and doubly represented quark flavours respectively.

We can also consider changing the momenta to access different kinematics. By
boosting the initial and final states while keeping the momentum transfer
constant, we can access smaller values of \(Q^2\).
We can also increase the three-momentum of the current insertion, giving
access to larger values of \(Q^2\).
For such kinematics at all masses we find that in general, the conventional
plateaus are later in time and take smaller values than the PEVA plateaus.

These results indicate that the PEVA technique is critical to the
correct extraction of the electric form factors of this nucleon excitation.
The conventional analysis is contaminated by opposite-parity states,
and when these states are removed by the PEVA technique it has a significant
effect on the extracted form factor values. Hence, we now focus our
attention only on the PEVA results for the remainder of this subsection.

\begin{table*}
    \caption{\label{tab:formfactors:kinematics}Different kinematics used in our
        analysis to access a range of \(Q^2\) values. The \(Q^2\) value listed
        is for the first negative-parity nucleon excitation at the middle pion mass of
        \(m_{\pi} = \SI{411}{\mega\electronvolt}\). The statistical error listed
        for \(Q^2\) comes from both the determination of the mass of the state
        and the conversion to physical units.
        In the so-called Breit frame kinematics, where the incoming and
        outgoing energies are equal, the correlated statistical errors from the mass
        cancel exactly, and as such the only source of errors is uncertainty in
        the lattice spacing used in converting to physical units.}
    \sisetup{%
    table-number-alignment = center,
    table-figures-integer = 1,
    table-figures-decimal = 0,
    table-figures-uncertainty = 0,
}
\begin{tabular*}{\linewidth}{%
@{\extracolsep{\fill}}
        c
        c
        c
        S[table-figures-integer = 1, table-figures-decimal = 4, table-figures-uncertainty = 3]
}
    \toprule{} \vspace{-9pt}\\
    {Source momentum \(\vect{p}\)} & {Sink momentum \(\vect{p}'\)} & {Momentum transfer \(\vect{q}\)} & {\(Q^2 \, / \, \si{\giga\electronvolt^2}\)} \\
    \colrule{} \vspace{-9pt}\\
    \((2,0,0)\) & \((3,0,0)\) & \((1,0,0)\) & 0.1224 \pm 0.0035 \\
    \((2,0,1)\) & \((3,0,1)\) & \((1,0,0)\) & 0.1239 \pm 0.0035 \\
    \((1,0,0)\) & \((2,0,0)\) & \((1,0,0)\) & 0.1454 \pm 0.0040 \\
    \((1,0,1)\) & \((2,0,1)\) & \((1,0,0)\) & 0.1462 \pm 0.0040 \\
    \((0,0,0)\) & \((1,0,0)\) & \((1,0,0)\) & 0.1604 \pm 0.0044 \\
    \((0,0,1)\) & \((1,0,1)\) & \((1,0,0)\) & 0.1606 \pm 0.0044 \\
    \((2,0,0)\) & \((3,1,0)\) & \((1,1,0)\) & 0.2683 \pm 0.0074 \\
    \((1,0,0)\) & \((2,1,0)\) & \((1,1,0)\) & 0.2953 \pm 0.0081 \\
    \((0,0,0)\) & \((1,1,0)\) & \((1,1,0)\) & 0.3169 \pm 0.0086 \\
    \((0,-1,0)\) & \((1,0,0)\) & \((1,1,0)\) & 0.3251 \pm 0.0089 \\
    \((1,0,0)\) & \((3,0,0)\) & \((2,0,0)\) & 0.5404 \pm 0.0150 \\
    \((0,0,0)\) & \((2,0,0)\) & \((2,0,0)\) & 0.6190 \pm 0.0169 \\
    \((-1,0,0)\) & \((1,0,0)\) & \((2,0,0)\) & 0.6502 \pm 0.0177 \\
    \botrule{}
\end{tabular*}
\end{table*}

\begin{figure}[tbp]
    {\centering
        \includegraphics{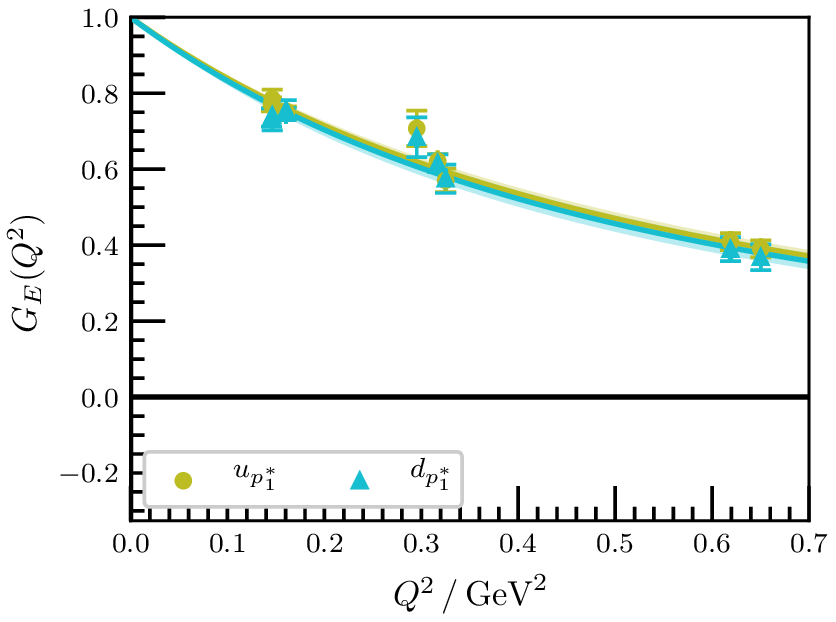}}
    \caption{\label{fig:1stneg:GE:k3Q2qs}Quark-flavour contributions to
        the electric form factor for the first negative-parity
        excitation at \(m_{\pi} = \SI{411}{\mega\electronvolt}\). The
        curves are dipole fits to the form factors, with the \(y\)-intercept
        fixed to unity.
        They correspond to RMS charge radii of \SI{0.654 \pm 0.020}{\femto\meter}
        for the doubly represented quark flavour (\(\indSdblsa\)) and
        \SI{0.670 \pm 0.026}{\femto\meter} for the singly represented quark flavour
        (\(\indSsingsa\)).
        }
\end{figure}

In Fig.~\ref{fig:1stneg:GE:k3Q2qs}, we plot the \(Q^2\) dependence of the
electric form factor at \(m_{\pi} = \SI{411}{\mega\electronvolt}\). The set of
kinematics used to access the various \(Q^2\) values is listed in
Table~\ref{tab:formfactors:kinematics}, and we exclude any fits for which there
is no acceptable plateau, or the variational analysis fails. We see that the two
quark flavours have very similar contributions to the electric form factor.
They both agree well with a dipole ansatz
\begin{equation}
    \ffDipole{} = \frac{\dipoleYIntercept{}}{{\left(1 + Q^2 / \dipoleScale{}^2\right)}^2}\,,
\end{equation}
with \(\dipoleYIntercept{}\) fixed to one, as we are working with single quarks of unit
charge. These fits correspond to a RMS charge radius\index{charge radius} of
\SI{0.654 \pm 0.020}{\femto\meter} for the doubly represented
quark flavour and \SI{0.670 \pm 0.026}{\femto\meter} for the singly represented
quark flavour. These charge radii are similar to the charge radii of the
individual quark sectors in the ground state examined in Ref.~\cite{Stokes:2018emx},
(\SI{0.662 \pm 0.012}{\femto\meter} for the doubly
represented quark flavour and \SI{0.633 \pm 0.012}{\femto\meter} for the singly
represented quark flavour).
The doubly represented quark sector agrees to within one standard deviation.
However, the singly represented quark sector in the excitation has a charge
radius\index{charge radius} approximately \num{1.5} standard deviations
larger than the ground state.

We see similar behaviour for the other four masses. In all cases, the quark
distributions are much smaller than the lattice length \(L\sim\SI{3}{\femto\meter}\).
The plots for these masses are omitted from this paper for the sake of brevity.

\subsubsection{Constituent quark model expectations}

Within the context of a simple constituent quark model, the near equivalence of the electric charge
radii of quark sectors within the ground state nucleon and the first negative parity
excitation seems truly remarkable.
Considering the effective potential of the radial Schr\"odinger equation, one expects the
repulsive centripetal term proportional to $\ell\, (\ell +1)$ to force the quarks to larger radii
for odd-parity $\ell = 1$ states.

However, one needs to recall that these radii are from quantum field theory where dynamical
quark-antiquark pairs enable the creation of meson-nucleon components in the $N^*$ states.  The
meson provides the negative parity such that all quarks can reside in relative $s$-waves within the
hadrons, forming an $S$-wave meson-baryon molecule.  In this way the centripetal barrier is avoided
and the negative-parity states can have a size similar to the ground state nucleon.
Relevant meson-baryon channels for the odd-parity states include \(K \Sigma\) and \(K \Lambda\) in
addition to the standard \(\eta N\) and \(\pi N\) meson-baryon channels.

\subsubsection{Baryon electric form factors}

In order to compute the form factors of the first negative-parity excitation of
the proton, \(\ffElectricSprsa\), and neutron, \(\ffElectricSnesa\), we need to
take the correct linear combinations of the contributions from the doubly
represented quark flavour and the singly represented quark flavour to
reintroduce the multiplicity of the doubly represented quark and the physical
charges of the up and down quarks. To this end we define
\begin{subequations}
\begin{align}
    \ffElectricSprsa{} &\definedby +\frac{4}{3}\ffElectricSdblsa{} - \frac{1}{3} \ffElectricSsingsa{}\,, \neweqn
    \ffElectricSnesa{} &\definedby -\frac{2}{3}\ffElectricSdblsa{} + \frac{2}{3} \ffElectricSsingsa{}\,.
\end{align}
\end{subequations}

\begin{figure}[tbp]
    {\centering
        \includegraphics{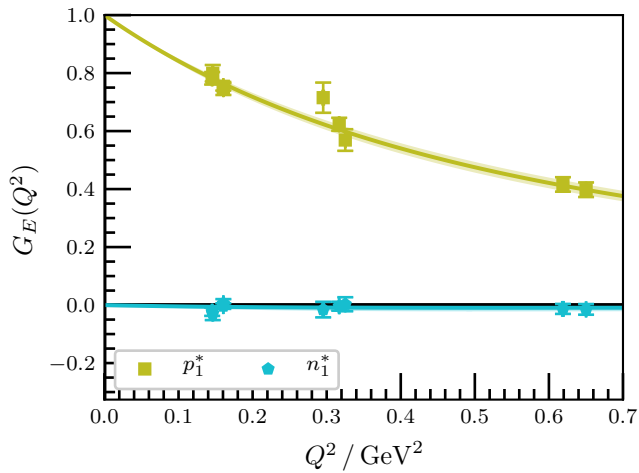}}
    \caption{\label{fig:1stneg:GE:k3Q2nucleon}\(\ffElectric\) for the
        first negative-parity excitations of the proton and
        neutron at \(m_{\pi} = \SI{411}{\mega\electronvolt}\).
        The curves correspond to linear combinations
        of the dipole fits to the individual quark sectors from
        Fig.~\ref{fig:1stneg:GE:k3Q2qs}. These combinations provide a squared
        charge radius of \SI{0.421 \pm 0.029}{\femto\meter\squared}
        for the proton and \SI{0.014 \pm 0.018}{\femto\meter\squared} for the neutron.
        }
\end{figure}

In Fig.~\ref{fig:1stneg:GE:k3Q2nucleon}, we
plot the nucleon electric form factors obtained by taking these combinations
of the form factors at \(m_{\pi} = \SI{411}{\mega\electronvolt}\). The form factor
for the neutron excitation is close to zero, reflecting the similar charge radii
of the individual quark flavours. By combining the dipole fits to the individual
quark sectors in the same way as the data points one obtains a model for the
\(Q^2\) dependence of the electric form factors of the excited proton and neutron
that includes full information from both quark sectors.
If we do this for all five pion masses, we extract squared charge
radii\index{charge radius} for the proton excitation ranging from
\SI{0.340 \pm 0.029}{\femto\meter\squared} to \SI{0.470 \pm 0.054}{\femto\meter\squared},
increasing with decreasing pion mass.
For the neutron excitation, the squared charge radii are close to or slightly below zero, for example
\SI{-0.033 \pm 0.024}{\femto\meter\squared} at
\(m_{\pi} = \SI{296}{\mega\electronvolt}\).

\begin{figure}[tbp]
    {\centering
        \includegraphics{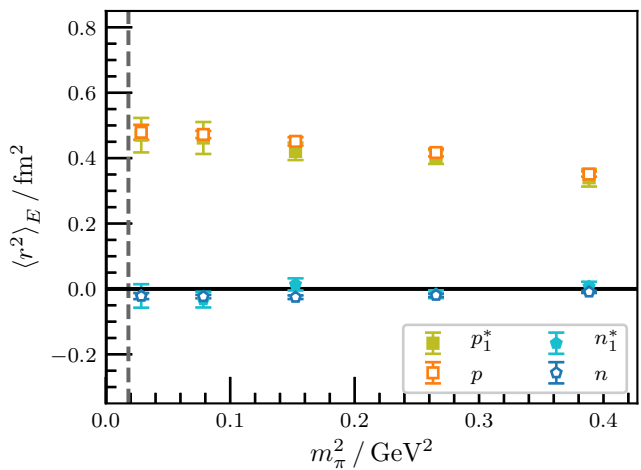}}
    \caption{\label{fig:1stneg:GE:proton}Quark-mass dependence of squared charge radii
    for the first negative-parity excitation of the proton and neutron.
    Results are obtained from dipole fits to the electric form factors of the individual
    quark sectors. For comparison, the radii for the ground states are
    plotted with open points. We see a clear trend to larger charge radii as the pion mass
    approaches the physical pion, represented by the dashed vertical line.
    }
\end{figure}

As illustrated in Fig.~\ref{fig:1stneg:GE:proton}, the pion-mass dependence is fairly smooth, and
has a clear trend to larger radii at lower pion masses. There is no hint of significant
non-analytic behaviour in the quark-mass dependence, due to finite-volume
suppression~\cite{Hall:2012yx,Hall:2013oga}.

For each pion mass considered, the extracted squared charge radii\index{charge radius} for this
first negative-parity excitation are consistent with the radii of the ground-state proton and
neutron at the same mass, as obtained in Ref.~\cite{Stokes:2018emx}.

As discussed in the context of the quark sector contributions, meson-nucleon components in the wave
function enable all quarks to reside in relative $s$-waves within the hadrons, forming an
$S$-wave meson-nucleon molecule.  In this way, the negative-parity states can have a size similar
to the ground state nucleon.

\subsection{\texorpdfstring{\(G_M\)}{GM} for first negative-parity excitation\label{sec:excitations:1stneg:GM}}

\subsubsection{Quark-flavour contributions}

\begin{figure}[tbp]
    {\centering
        \includegraphics{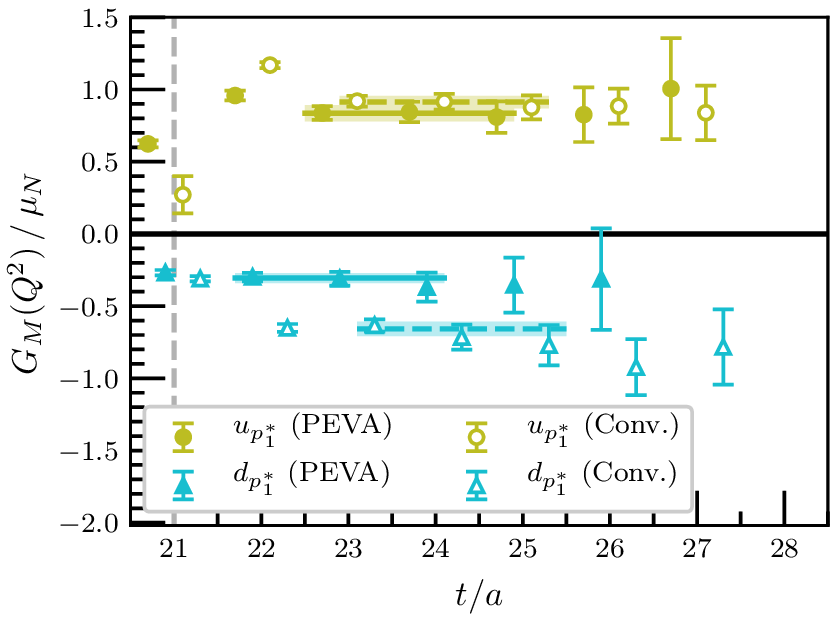}}
    \caption{\label{fig:1stneg:GM:k3p000pp100}Quark-flavour contributions to
        \(\ffMagnetic\) for the first
        negative-parity excitation at \(m_{\pi} = \SI{411}{\mega\electronvolt}\)
        for the lowest-momentum kinematics, providing
        \(Q^2 = \SI{0.1604 \pm 0.0044}{\giga\electronvolt^2}\).         As in Fig.~\ref{fig:1stneg:GE:k1p000pp100}, we plot the conventional
        analysis with open markers and dashed fit lines
        and the new PEVA approach with filled markers and solid fit lines.
        For the doubly represented quark flavour, the plateaus for both
        analyses are from \num{23}--\num{25} and take consistent values. For
        the singly represented quark flavour, the PEVA fit is
        from \num{22}--\num{24}, while the conventional fit is from
        \num{23}--\num{25}, and has a significantly more negative value.}
                                \end{figure}

We now proceed to the magnetic form factor. In Fig.~\ref{fig:1stneg:GM:k3p000pp100},
we plot the plateaus at \(m_{\pi} = \SI{411}{\mega\electronvolt}\) with the
lowest-momentum kinematics. Here we present results in terms of nuclear
magnetons, \(\mu_N \definedby \frac{e \hbar}{2 \massPhysSpr{}}\), defined in
terms of the physical proton mass, \(\massPhysSpr{}\). While
the conventional and PEVA plateaus for the doubly represented quark flavour are
consistent, both in fit region and value, the conventional plateau for the
singly represented quark flavour starts later and has a significantly
more negative value than the PEVA plateau. We see a similar effect at all five
pion masses and a variety of kinematics.

\begin{figure}[tbp]
    {\centering
        \includegraphics{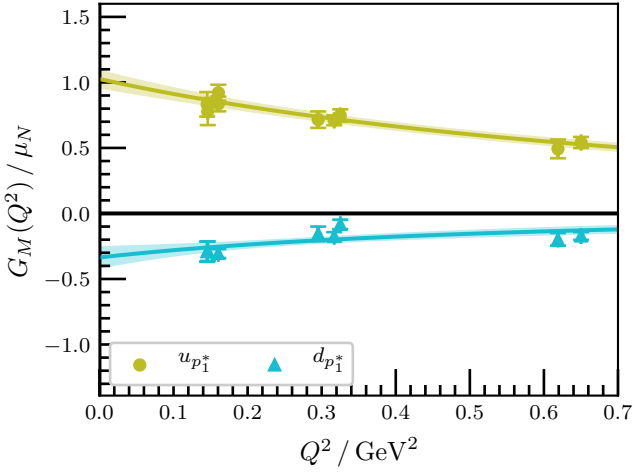}}
    \caption{\label{fig:1stneg:GM:k3Q2qs}Quark-flavour contributions to
        \(\ffMagnetic\) for the first
        negative-parity excitation at \(m_{\pi} = \SI{411}{\mega\electronvolt}\).
        The curves are dipole fits to the individual quark sectors.
                        }
\end{figure}

Having fit the form factor plateaus, we can investigate the \(Q^2\) dependence of \(\ffMagnetic\).
In Fig.~\ref{fig:1stneg:GM:k3Q2qs}, we plot the contributions to \(\ffMagnetic\)
from both the singly represented quark flavour and the doubly represented
quark flavour at \(m_{\pi} = \SI{411}{\mega\electronvolt}\). Both
quark flavours are consistent with a dipole fit. The \(Q^2\) dependence is similar to that for \(\ffElectric\), and the same is true
for the other pion masses considered.

\subsubsection{Baryon magnetic form factors}

\begin{figure}[tbp]
    {\centering
        \includegraphics{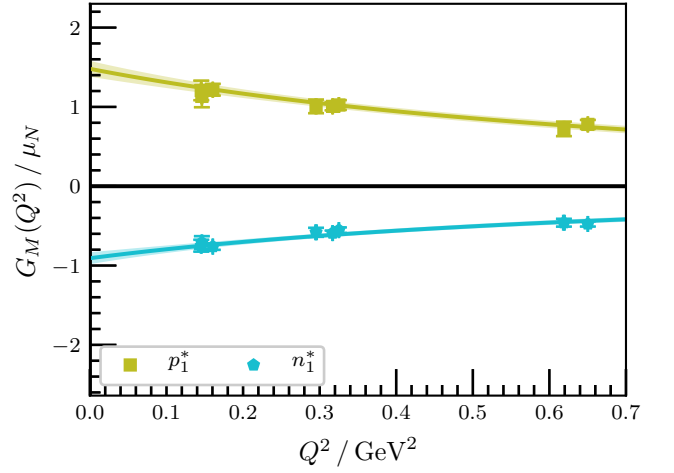}}
    \caption{\label{fig:1stneg:GM:k3Q2nucleon}\(\ffMagnetic\) for the
        first negative-parity excitations of the proton and neutron
        at \(m_{\pi} = \SI{411}{\mega\electronvolt}\).
        The curves correspond to linear combinations of the
        quark-sector dipole fits from Fig.~\ref{fig:1stneg:GM:k3Q2qs}.
                        }
\end{figure}

As described in Section~\ref{sec:excitations:1stneg:GE}, we can take
linear combinations of the individual quark
flavour contributions to compute the magnetic form factors of the excited
proton and neutron.
We plot these combinations for \(m_{\pi} = \SI{411}{\mega\electronvolt}\) in
Fig.~\ref{fig:1stneg:GM:k3Q2nucleon}. The squared magnetic radii given by
\begin{equation}
    \frac{\rsqMagnetic{}}{\mm{}} = \frac{-6}{\ffMagnetic[0]{}} \, {\left.\frac{d\ffMagnetic{}}{dQ^2}\right|}_{Q^2 = 0}\,,
\end{equation}
are obtained via dipole fits to the quark sector contributions, which allow the
value and slope of \(\ffMagnetic{}\) to be extrapolated to \(Q^2 = 0\).
In obtaining hadronic magnetic radii, the quark sectors combine with additional
weightings given by \(\ffMagnetic[0]{}\), that is
\begin{subequations}
\begin{align}
    \frac{\rsqMagneticSprsa{}}{\mmSprsa{}} &\definedby \frac{1}{+\frac{4}{3}\ffMagneticSdblsa[0]{} - \frac{1}{3}\ffMagneticSsingsa[0]{}} \conteqn
    &\!\!\!\!\!\!\!\!\!\!\!\times\!\left(+\frac{4}{3}\ffMagneticSdblsa[0]{}\frac{\rsqMagneticSdblsa{}}{\mmSdblsa{}}
    - \frac{1}{3} \ffMagneticSsingsa[0]{} \frac{\rsqMagneticSsingsa{}}{\mmSsingsa{}}\right), \neweqn
    \frac{\rsqMagneticSnesa{}}{\mmSnesa{}} &\definedby \frac{1}{-\frac{2}{3}\ffMagneticSdblsa[0]{} + \frac{2}{3}\ffMagneticSsingsa[0]{}} \conteqn
    &\!\!\!\!\!\!\!\!\!\!\!\times\!\left(-\frac{2}{3}\ffMagneticSdblsa[0]{}\frac{\rsqMagneticSdblsa{}}{\mmSdblsa{}}
    + \frac{2}{3} \ffMagneticSsingsa[0]{} \frac{\rsqMagneticSsingsa{}}{\mmSsingsa{}}\right).
\end{align}
\end{subequations}
For all five masses, we find that these squared magnetic radii mostly agree
with the charge radii from \(\ffElectric\).

\begin{figure}[tbp]
    {\centering
        \includegraphics{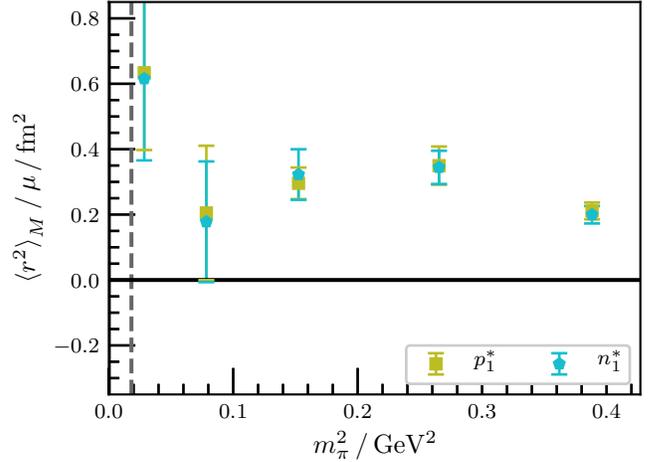}}
    \caption{\label{fig:1stneg:GM:radius}Quark-mass dependence of
        squared magnetic radii for the first negative-parity excitation of the proton.
        }
\end{figure}

In Fig.~\ref{fig:1stneg:GM:radius}, we plot the pion-mass dependence of the
squared magnetic radius obtained from quark-sector dipole fits to \(\ffMagnetic\) for
the excited proton and neutron.

\subsubsection{Baryon magnetic moments}

Returning to the individual quark sector results, we note that \(\ffMagnetic\)
and \(\ffElectric\) have a similar \(Q^2\) dependence over the range considered.
In light of this, we hypothesise that \(\ffMagnetic{}\) and
\(\ffElectric{}\) have the same \(Q^2\) scaling in this region.
If this hypothesis is valid, then the ratio of \(\ffMagnetic{}\) to
\(\ffElectric{}\) should be independent of \(Q^2\). Since we are working with
an improved conserved vector current, and single quarks of unit charge,
\(\ffElectric[0]{} = 1\) exactly, and \(\ffMagnetic[0]{}\) is the
contribution of the quark flavour
to the magnetic moment (up to scaling by the physical charge). Hence, the ratio
\begin{equation}
    \mmEff{}(Q^2) \definedby \frac{\ffMagnetic{}}{\ffElectric{}}\,,
\end{equation}
is expected to provide a measure of the contribution to the magnetic
moment from the given quark flavour.

\begin{figure}[tbp]
    {\centering
        \includegraphics{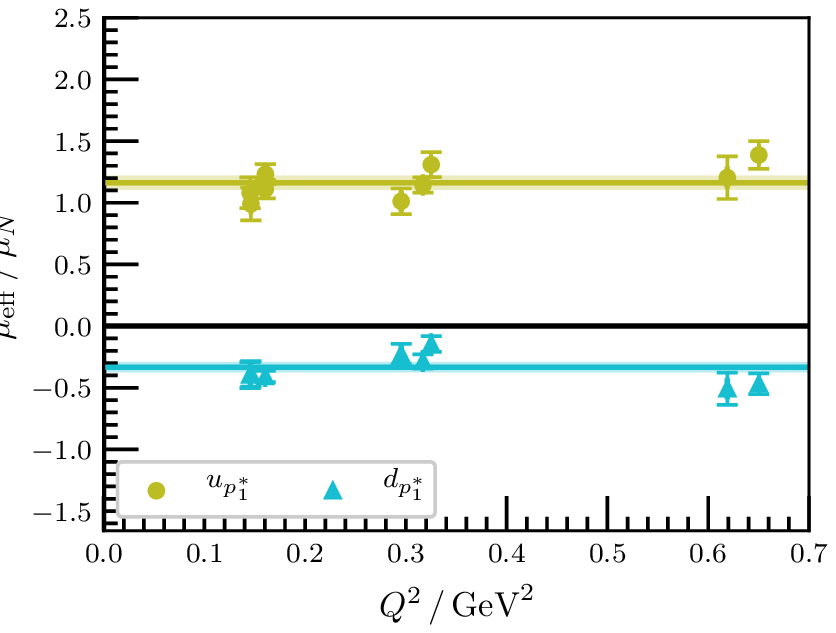}}
    \caption{\label{fig:1stneg:mm:k3}\(\mmEff{}\) for individual quarks of unit
        charge in the first negative-parity nucleon excitation at
        \(m_{\pi} = \SI{411}{\mega\electronvolt}\). The shaded bands are
        constant fits to the effective magnetic moment, corresponding to magnetic moment contributions
        of \num{1.163 \pm 0.060} \(\mu_N\) for the doubly represented quark and
        \num{-0.333 \pm 0.043} \(\mu_N\) for the singly represented quark.
        }
\end{figure}

In Fig.~\ref{fig:1stneg:mm:k3}, we plot this ratio at
\(m_{\pi} = \SI{411}{\mega\electronvolt}\) as a function of \(Q^2\). We see
that as expected, the ratio is approximately constant across the
\(Q^2\) range accessible by our kinematics. This holds true for all five
pion masses considered in this work.
This supports the underlying hypothesis that the
\(Q^2\) scaling of the contributions to \(\ffElectric\) and \(\ffMagnetic\)
from each quark sector is
the same, and hence suggests that \(\mmEff{}\) is a good estimate for the
magnetic moment\index{magnetic moment} of this state.

\begin{figure}[tbp]
    {\centering
        \includegraphics{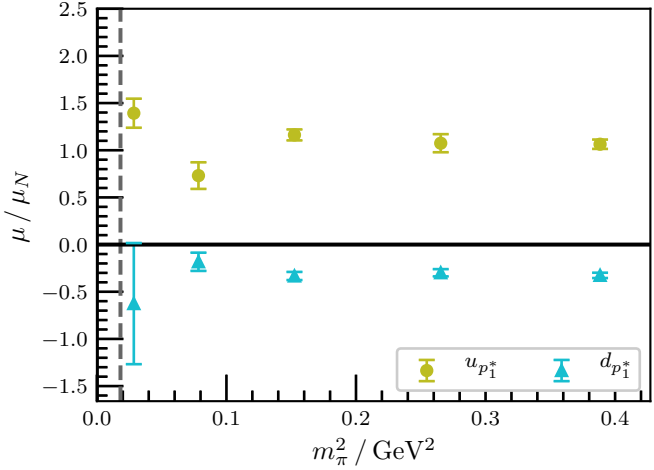}}
    \caption{\label{fig:1stneg:mm:qs}Quark-mass dependence of contributions
        from individual unit-charge quarks to the magnetic moment of the first
        negative-parity nucleon excitation. The vertical
        dashed line corresponds to the physical pion mass.}
\end{figure}

\begin{figure}[tbp]
    {\centering
        \includegraphics{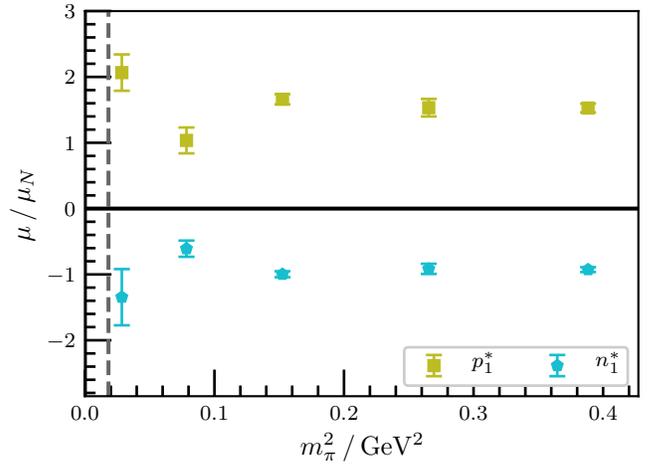}}
    \caption{\label{fig:1stneg:mm:nucleon}Quark-mass dependence of
        the magnetic moment of the first negative-parity excitations of the
        proton and neutron.}
\end{figure}

We take constant fits to \(\mmEff\) at each quark mass, and plot their pion-mass
dependence in Fig.~\ref{fig:1stneg:mm:qs}. By taking linear
combinations of these fits as described for \(\ffElectric\) and \(\ffMagnetic\)
above, we obtain magnetic moment\index{magnetic moment} estimates for the excited proton and neutron,
as plotted in Fig.~\ref{fig:1stneg:mm:nucleon}.
For the heaviest three pion masses, the effective
magnetic moments show little pion mass dependence and have tight error bars.
The lightest two pion masses have much larger errors, and we observe a
discontinuity in \(\ffMagnetic\) at the the second lightest mass, appearing
as a significantly smaller magnetic moment for both states.
This suggests that there could be
a change in the structure of this state at that mass. However, there is no
corresponding change in \(\ffElectric{}\).
At the lightest mass, the magnetic moments appear to return to consistency with
the values from the heavier masses.
Hence, it is unclear whether the behaviour at the second
lightest mass indicates a change in the nature of the state, the presence of
significant scattering-state contamination, or is a
result of increasing gauge noise at lighter pion masses.
Further research will be required to
determine which of these possibilities is realised.

In this section focusing on the first negative-parity excitation of the nucleon,
we have demonstrated the importance of the PEVA technique in correctly extracting
both the electric and magnetic form factors. From these, we derived charge radii,
magnetic moments and magnetic radii.

While we regard the results for the three heaviest quark masses to be robust,
we take some caution with the lightest quark mass results as they may be
influenced by important unaccounted scattering-state contributions.

\begin{table}[hbt]
    \caption{\label{tab:formfactors:state0}Radii and magnetic
    moments of the ground state proton and neutron. Radii are obtained from from
    combinations of quark-sector dipole fits and magnetic moments are obtained
    from quark-sector ratios of \(\ffMagnetic\) to \(\ffElectric{}\).}
    \sisetup{%
    table-number-alignment = center,
    table-figures-integer = 1,
    table-figures-decimal = 0,
    table-figures-uncertainty = 0,
}
\begin{tabular*}{\linewidth}{%
@{\extracolsep{\fill}}
        S[table-figures-integer = 1, table-figures-decimal = 4, table-figures-uncertainty = 3]
        S[table-figures-integer = 1, table-figures-decimal = 3, table-figures-uncertainty = 2]
        S[table-figures-integer = 1, table-figures-decimal = 3, table-figures-uncertainty = 2]
        S[table-figures-integer = 1, table-figures-decimal = 2, table-figures-uncertainty = 1]
}
    \toprule{} \vspace{-9pt}\\
    {\(m_{\pi}^2 \, / \, \si{\giga\electronvolt^2}\)} & {\(\rsqElectricSpr{} \, / \, \si{\femto\meter\squared}\)} & {\(\rsqMagneticSpr{} \, / \, \mmSpr{}\! \, / \, \si{\femto\meter\squared}\)} & {\(\mmSpr{}\! \, / \, \mu_N\)} \\
    \colrule{} \vspace{-9pt}\\
    0.3884 \pm 0.0113 & 0.351 \pm 0.013 & 0.301 \pm 0.013 & 1.89 \pm 0.03 \\
    0.2654 \pm 0.0081 & 0.417 \pm 0.017 & 0.341 \pm 0.015 & 2.10 \pm 0.04 \\
    0.1525 \pm 0.0043 & 0.451 \pm 0.018 & 0.340 \pm 0.013 & 2.24 \pm 0.04 \\
    0.0784 \pm 0.0025 & 0.472 \pm 0.017 & 0.360 \pm 0.016 & 2.30 \pm 0.04 \\
    0.0285 \pm 0.0012 & 0.479 \pm 0.027 & 0.324 \pm 0.036 & 2.52 \pm 0.08 \\
    \botrule{}
\end{tabular*}
    \sisetup{%
    table-number-alignment = center,
    table-figures-integer = 1,
    table-figures-decimal = 0,
    table-figures-uncertainty = 0,
}
\begin{tabular*}{\linewidth}{%
@{\extracolsep{\fill}}
        S[table-figures-integer = 1, table-figures-decimal = 4, table-figures-uncertainty = 3]
        S[table-sign-mantissa,table-figures-integer = 2, table-figures-decimal = 3, table-figures-uncertainty = 1]
        S[table-figures-integer = 1, table-figures-decimal = 3, table-figures-uncertainty = 2]
        S[table-sign-mantissa,table-figures-integer = 2, table-figures-decimal = 2, table-figures-uncertainty = 1]
}
    \toprule{} \vspace{-9pt}\\
    {\(m_{\pi}^2 \, / \, \si{\giga\electronvolt^2}\)} & {\(\rsqElectricSne{} \, / \, \si{\femto\meter\squared}\)} & {\(\rsqMagneticSne{} \, / \, \mmSne{}\! \, / \, \si{\femto\meter\squared}\)} & {\(\mmSne{}\! \, / \, \mu_N\)} \\
    \colrule{} \vspace{-9pt}\\
    0.3884 \pm 0.0113 & -0.009 \pm 0.003 & 0.303 \pm 0.013 & -1.20 \pm 0.02 \\
    0.2654 \pm 0.0081 & -0.019 \pm 0.005 & 0.337 \pm 0.015 & -1.33 \pm 0.03 \\
    0.1525 \pm 0.0043 & -0.025 \pm 0.006 & 0.350 \pm 0.014 & -1.38 \pm 0.03 \\
    0.0784 \pm 0.0025 & -0.023 \pm 0.005 & 0.376 \pm 0.020 & -1.40 \pm 0.03 \\
    0.0285 \pm 0.0012 & -0.022 \pm 0.009 & 0.437 \pm 0.084 & -1.57 \pm 0.05 \\
    \botrule{}
\end{tabular*}
\end{table}

\begin{table}[hbt]
    \caption{\label{tab:formfactors:state1}Radii and magnetic
    moments of the first negative parity excitation of the proton and neutron.
    Radii are obtained from from combinations of quark-sector dipole fits and magnetic moments are obtained
    from quark-sector ratios of \(\ffMagnetic\) to \(\ffElectric{}\).}
    \sisetup{%
    table-number-alignment = center,
    table-figures-integer = 1,
    table-figures-decimal = 0,
    table-figures-uncertainty = 0,
}
\begin{tabular*}{\linewidth}{%
@{\extracolsep{\fill}}
        S[table-figures-integer = 1, table-figures-decimal = 4, table-figures-uncertainty = 3]
        S[table-figures-integer = 1, table-figures-decimal = 3, table-figures-uncertainty = 2]
        S[table-figures-integer = 1, table-figures-decimal = 2, table-figures-uncertainty = 2]
        S[table-figures-integer = 1, table-figures-decimal = 2, table-figures-uncertainty = 2]
}
    \toprule{} \vspace{-9pt}\\
    {\(m_{\pi}^2 \, / \, \si{\giga\electronvolt^2}\)} & {\(\rsqElectricSprsa{} \, / \, \si{\femto\meter\squared}\)} & {\(\rsqMagneticSprsa{} \, / \, \mmSprsa{}\! \, / \, \si{\femto\meter\squared}\)} & {\(\mmSprsa{}\! \, / \, \mu_N\)} \\
    \colrule{} \vspace{-9pt}\\
    0.3884 \pm 0.0113 & 0.340 \pm 0.029 & 0.21 \pm 0.03 & 1.53 \pm 0.07 \\
    0.2654 \pm 0.0081 & 0.403 \pm 0.024 & 0.35 \pm 0.06 & 1.53 \pm 0.14 \\
    0.1525 \pm 0.0043 & 0.421 \pm 0.029 & 0.30 \pm 0.05 & 1.66 \pm 0.08 \\
    0.0784 \pm 0.0025 & 0.462 \pm 0.050 & 0.21 \pm 0.21 & 1.04 \pm 0.20 \\
    0.0285 \pm 0.0012 & 0.470 \pm 0.054 & 0.63 \pm 0.24 & 2.07 \pm 0.28 \\
    \botrule{}
\end{tabular*}
    \sisetup{%
    table-number-alignment = center,
    table-figures-integer = 1,
    table-figures-decimal = 0,
    table-figures-uncertainty = 0,
}
\begin{tabular*}{\linewidth}{%
@{\extracolsep{\fill}}
        S[table-figures-integer = 1, table-figures-decimal = 4, table-figures-uncertainty = 3]
        S[table-sign-mantissa,table-figures-integer = 2, table-figures-decimal = 3, table-figures-uncertainty = 2]
        S[table-figures-integer = 1, table-figures-decimal = 2, table-figures-uncertainty = 2]
        S[table-sign-mantissa,table-figures-integer = 2, table-figures-decimal = 2, table-figures-uncertainty = 2]
}
    \toprule{} \vspace{-9pt}\\
    {\(m_{\pi}^2 \, / \, \si{\giga\electronvolt^2}\)} & {\(\rsqElectricSnesa{} \, / \, \si{\femto\meter\squared}\)} & {\(\rsqMagneticSnesa{} \, / \, \mmSnesa{}\! \, / \, \si{\femto\meter\squared}\)} & {\(\mmSnesa{}\! \, / \, \mu_N\)} \\
    \colrule{} \vspace{-9pt}\\
    0.3884 \pm 0.0113 & 0.007 \pm 0.015 & 0.20 \pm 0.03 & -0.93 \pm 0.04 \\
    0.2654 \pm 0.0081 & -0.016 \pm 0.012 & 0.34 \pm 0.05 & -0.92 \pm 0.08 \\
    0.1525 \pm 0.0043 & 0.014 \pm 0.018 & 0.32 \pm 0.08 & -1.00 \pm 0.05 \\
    0.0784 \pm 0.0025 & -0.033 \pm 0.024 & 0.18 \pm 0.18 & -0.61 \pm 0.12 \\
    0.0285 \pm 0.0012 & -0.021 \pm 0.036 & 0.62 \pm 0.25 & -1.35 \pm 0.43 \\
    \botrule{}
\end{tabular*}
\end{table}

In Table~\ref{tab:formfactors:state0},
we present the charge radii, magnetic radii, and magnetic moments of the ground state proton and
neutron. These measurements are obtained from the electric and magnetic form factors
as described above. We provide these results here for easy comparison with the
excited state results presented in this paper.

In Table~\ref{tab:formfactors:state1},
we present the same results for the first negative-parity excitation. We see
that this state has radii similar to the ground state, but notably different
magnetic moments. We find that at the heavier quark masses, these magnetic
moments agree well with constituent quark model predictions, as discussed
below in Section~\ref{sec:excitations:model}.

\subsection{\texorpdfstring{\(G_E\)}{GE} for the second negative-parity excitation\label{sec:excitations:2ndneg:GE}}

\subsubsection{Quark-flavour contributions}

\begin{figure}[tbp]
    {\centering
        \includegraphics{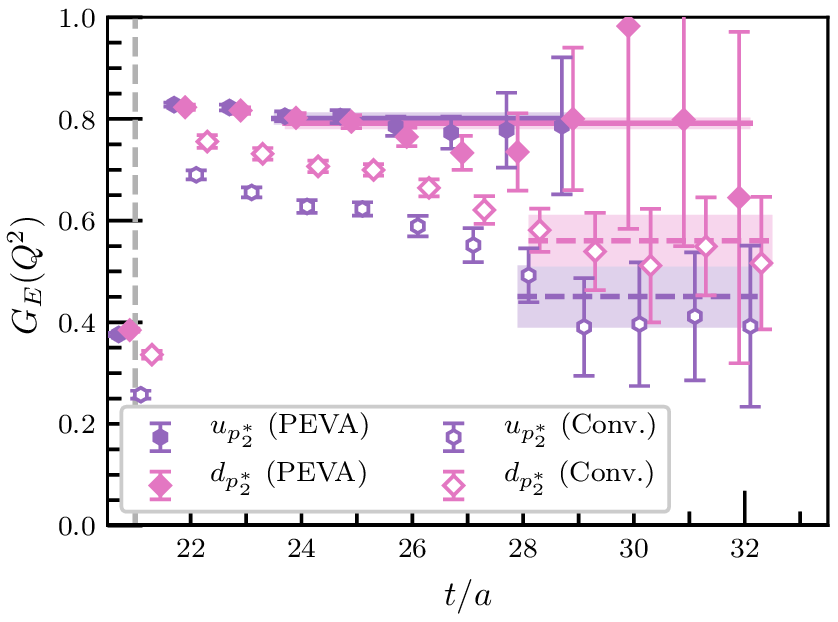}}
    \caption{\label{fig:2ndneg:GE:k1p000pp100}Quark-flavour contributions to
        the electric form factor for the second negative-parity nucleon
        excitation at \(m_{\pi} = \SI{702}{\mega\electronvolt}\) for the
        lowest-momentum kinematics, providing
        \(Q^2 = \SI{0.1425 \pm 0.0041}{\giga\electronvolt^2}\).         We plot the conventional analysis with open markers and dashed fit lines
        and the new PEVA approach with filled markers and solid fit lines.
        Results for both the singly represented quark flavour (\(\indSsingsb\))
        and the doubly represented quark flavour (\(\indSdblsb\)) are shown
        for single quarks of unit charge. Both PEVA fits are from time
        slice \num{24}, whereas the conventional fits both start at time slice
        \num{28}. The values for both conventional fits are significantly lower
        than the corresponding PEVA fits.
        }
\end{figure}

We now proceed to examine the next negative-parity excitation observed in this
study, \(\indSnsb\).
In Fig.~\ref{fig:2ndneg:GE:k1p000pp100}, we plot \(\ffElectric{}\) as a function
of sink time for both quark flavours at \(m_{\pi} = \SI{702}{\mega\electronvolt}\)
with the lowest-momentum kinematics. We see that the conventional extraction
sits even further below the PEVA extraction than the first negative-parity
excitation. While the PEVA fits start at time slice \num{24}, the conventional
fits are forced all the way out to time slice \num{28} and sit at only just above
half of the values of the PEVA fits.

\begin{figure}[tbp]
    {\centering
        \includegraphics{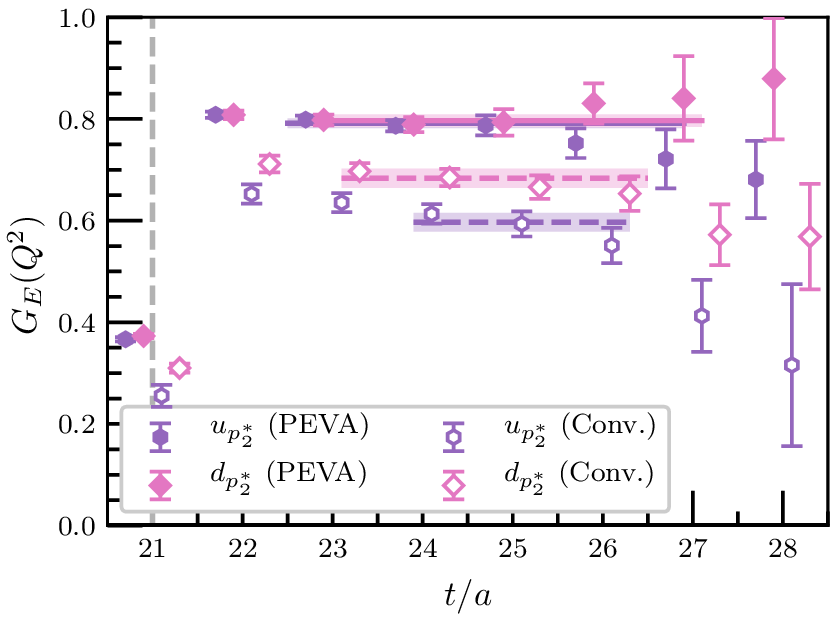}}
    \caption{\label{fig:2ndneg:GE:k2p000pp100}Quark-flavour contributions to
        \(\ffElectric\) for the second
        negative-parity excitation at \(m_{\pi} = \SI{570}{\mega\electronvolt}\)
        for the lowest-momentum kinematics, providing
        \(Q^2 = \SI{0.1458 \pm 0.0044}{\giga\electronvolt^2}\).         The conventions used in this plot are the same as in
        Fig.~\ref{fig:2ndneg:GE:k1p000pp100}.
        The conventional fits
        have significantly lower values than the PEVA fits, and the plateau for
        the doubly represented quark flavour starts one time slice later than the
        corresponding PEVA plateau.
        }
\end{figure}

Moving on to the lighter pion masses, the discrepancy between the extracted
form factors continues at \(m_{\pi} = \SI{570}{\mega\electronvolt}\) and
\(\SI{411}{\mega\electronvolt}\), with the conventional analysis giving
consistently lower values than PEVA\@. For example, Fig.~\ref{fig:2ndneg:GE:k2p000pp100}
shows the plateaus at \(m_{\pi} = \SI{570}{\mega\electronvolt}\).

At the lightest two pion masses, the signal gets significantly noisier,
and the difference between the two techniques gets harder to distinguish.
Increased statistics are required in order to clearly identify the effects of
opposite-parity contaminations of this state at these masses. However, in
principle, the enhancement of relativistic components of the baryon spinors at
light quark masses is expected to increase parity mixing in the conventional
analysis.

In addition, in the tail of the two point correlation function at the lightest
mass, contributions from low-lying states are
evident from a \(\chi^2\) analysis of a single state ansatz. This effect was
also seen in Ref.~\cite{Mahbub:2013bba}, where it was found that so long as a
single-state ansatz is satisfied in the fit region, this
contamination does not have a significant effect on the extracted mass. However,
as shown in Ref.~\cite{Stokes:2018emx}, contaminants that do not significantly
perturb the extracted mass can still have a significant effect on the extracted
form factors. As such, we must be cautious when interpreting results from this
state at this mass. To gain a deeper insight into this state at this mass,
multi-particle scattering operators will be necessary to properly isolate the
low-lying scattering state.

Returning to the other ensembles, we find that the trends from the lowest-momentum
kinematics continue for all other kinematics: for all masses for which the noise is
sufficiently low, the conventional fits sit significantly lower than the
PEVA fits.

Once again, these results clearly indicate that PEVA is critical to the correct
extraction of the electric form factors of this nucleon excitation. The
opposite-parity contaminations admitted by the conventional analysis
lead to significant underestimation of the value of the electric form
factor. Hence, we now focus our attention only on the PEVA results.

\begin{figure}[tbp]
    {\centering
        \includegraphics{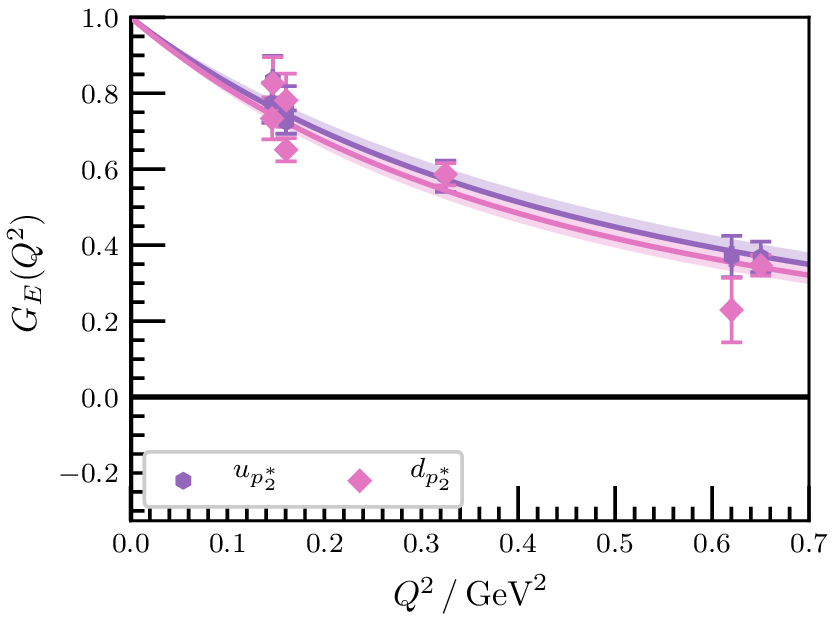}}
    \caption{\label{fig:2ndneg:GE:k3Q2qs}Quark-flavour contributions to
        \(\ffElectric\) for the second negative-parity excitation
        at \(m_{\pi} = \SI{411}{\mega\electronvolt}\). The
        curves are dipole fits to the form factor, with the \(y\)-intercept
        fixed to unity.
        They correspond to RMS charge
        radii of \SI{0.679 \pm 0.038}{\femto\meter} for the doubly represented quark
        (\(\indSdbl\)) and \SI{0.715 \pm 0.031}{\femto\meter} for the singly
        represented quark (\(\indSsing\)).
        }
\end{figure}

Plotting the acceptable plateaus as a function of \(Q^2\) reveals that the
contributions from the two quark flavours are once again very similar, and
agree well with a dipole
ansatz. For example, Fig.~\ref{fig:2ndneg:GE:k3Q2qs} shows dipole fits to the
two quark flavours at \(m_{\pi} = \SI{411}{\mega\electronvolt}\), with a RMS charge
radius\index{charge radius} of \SI{0.679 \pm 0.038}{\femto\meter} for the doubly represented quark flavour
and \SI{0.715 \pm 0.031}{\femto\meter} for the singly represented quark flavour.

The charge radius of the doubly represented quark flavour of \SI{0.679 \pm 0.038}{\femto\meter} is
similar to that seen in the ground state, \SI{0.662 \pm 0.012}{\femto\meter} \cite{Stokes:2018emx}
again suggesting a role for meson-baryon contributions escaping the centripetal barrier encountered
in a three-quark composition.  This time the RMS charge radius of the singly represented quark flavour of
\SI{0.715 \pm 0.031}{\femto\meter} is larger that that observed in the ground state, \SI{0.633
  \pm 0.012}{\femto\meter}.  However, it does overlap with that of the first negative parity
excitation at the one-sigma level.

\begin{figure}[tbp]
    {\centering
        \includegraphics{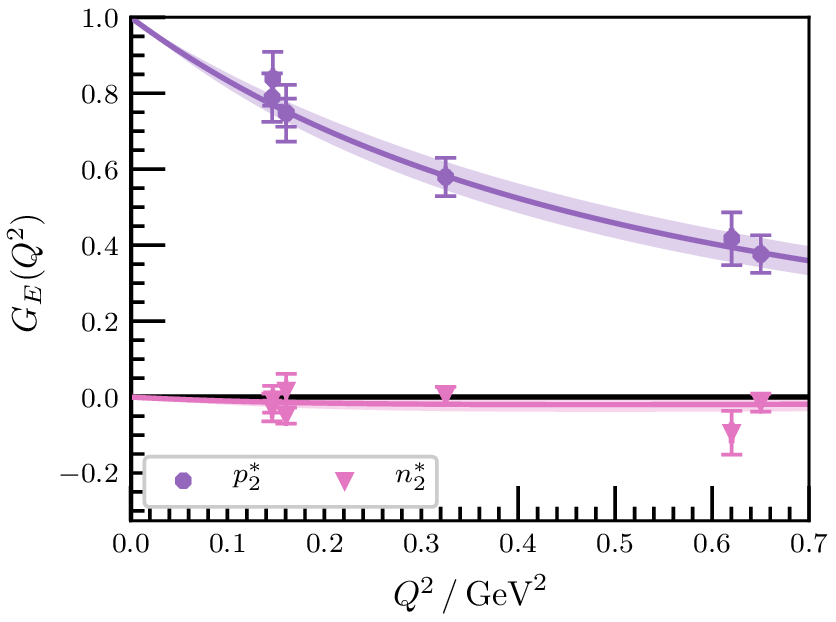}}
    \caption{\label{fig:2ndneg:GE:k3Q2nucleon}\(\ffElectric\) for the
        second negative-parity excitation of the proton and neutron
        at \(m_{\pi} = \SI{411}{\mega\electronvolt}\).
        The curves correspond to combinations of the
        quark-sector dipole fits from Fig.~\ref{fig:2ndneg:GE:k3Q2qs}, giving
        squared charge radii of \SI{0.445 \pm 0.062}{\femto\meter\squared}
        for the excited proton and \SI{0.033 \pm 0.030}{\femto\meter\squared} for the excited neutron.
        }
\end{figure}

\subsubsection{Baryon electric form factors}

We once again take the linear combinations discussed in
Section~\ref{sec:excitations:1stneg:GE} to form the excited proton
and neutron. For example, in Fig.~\ref{fig:2ndneg:GE:k3Q2nucleon},
we plot the electric form factors obtained from the quark-sector combinations at
\(m_{\pi} = \SI{411}{\mega\electronvolt}\).
At all five masses, the electric form factor
for the second negative-parity excitation of the neutron is approximately zero.

\begin{figure}[tbp]
    {\centering
        \includegraphics{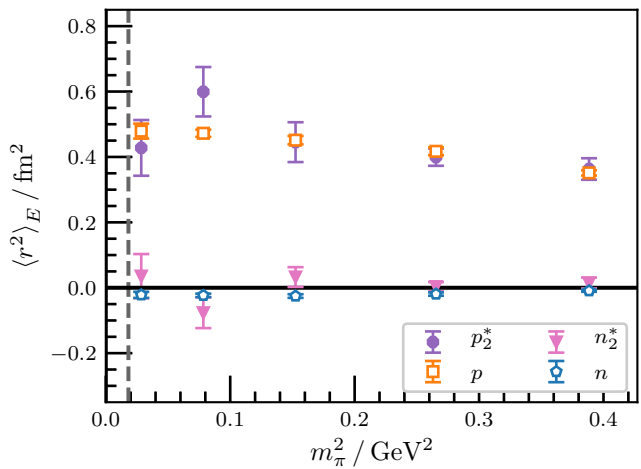}}
    \caption{\label{fig:2ndneg:GE:proton}Quark-mass dependence of charge radii
    from combinations of quark-sector dipole fits to \(\ffElectric\) for the second negative-parity
    excitation of the proton and neutron. The vertical dashed line corresponds to the physical pion
    mass, and the ground state is plotted with open points for comparison.}
\end{figure}

The pion-mass dependence of the charge radii is illustrated in Fig.~\ref{fig:2ndneg:GE:proton}.  At
the heaviest three pion masses, the linear combinations of the quark-sector dipole fits once again
provide charge radii consistent with the ground-state nucleon at the same quark
mass~\cite{Stokes:2018emx}, pointing to a role for meson-baryon $S$-wave Fock-space components,
escaping the centripetal barrier encountered in a three-quark composition.  We see that the
pion-mass dependence of the proton radius is fairly smooth at these heaviest masses, and has a
clear trend to increasing charge radius at lower pion masses.

\subsection{\texorpdfstring{\(G_M\)}{GM} for the second negative-parity excitation\label{sec:excitations:2ndneg:GM}}

\subsubsection{Quark-flavour contributions}

\begin{figure}[tbp]
    {\centering
        \includegraphics{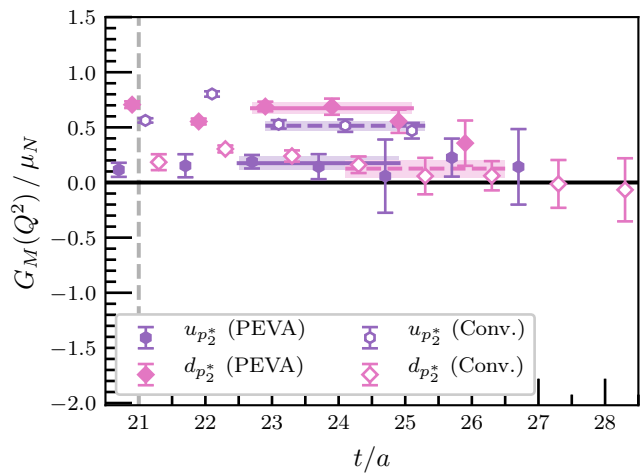}}
    \caption{\label{fig:2ndneg:GM:k1p000pp100}Quark-flavour contributions to
        \(\ffMagnetic\) for the second negative-parity excitation of the
        nucleon at \(m_{\pi} = \SI{702}{\mega\electronvolt}\) for the
        lowest-momentum kinematics, providing \(Q^2 = \SI{0.1425 \pm 0.0041}{\giga\electronvolt^2}\).         The conventions used in this plot are the same as in
        Fig.~\ref{fig:2ndneg:GE:k1p000pp100}.
        The plateaus for the PEVA technique both start at time slice \num{23}.
        The plateaus for the conventional analysis start at time slice \num{23}
        for \(\indSdblsb\) and time slice \num{24} for \(\indSsingsb\).
        The difference in the plateau values between the two analyses is enough
        to change the ordering of the two quark flavours.
        }
\end{figure}

We now advance to the magnetic form factor of this state. In
Fig.~\ref{fig:2ndneg:GM:k1p000pp100}, we plot the heaviest
pion mass of \(m_{\pi} = \SI{702}{\mega\electronvolt}\) and the lowest-momentum
kinematics. While the plateau time-regions for
the PEVA and conventional analysis are consistent, the values of those plateaus are
very different, and in fact change ordering between the two extractions.
We see a similar effect at \(m_{\pi} = \SI{296}{\mega\electronvolt}\),
\SI{411}{\mega\electronvolt}, and \SI{570}{\mega\electronvolt},
with similar inversions of the magnetic form factors between the two analyses.
We see similar
patterns for the other kinematics, with significantly different plateau values
between the two analyses when the statistical noise is low enough to distinguish them.

\begin{figure}[tbp]
    {\centering
        \includegraphics{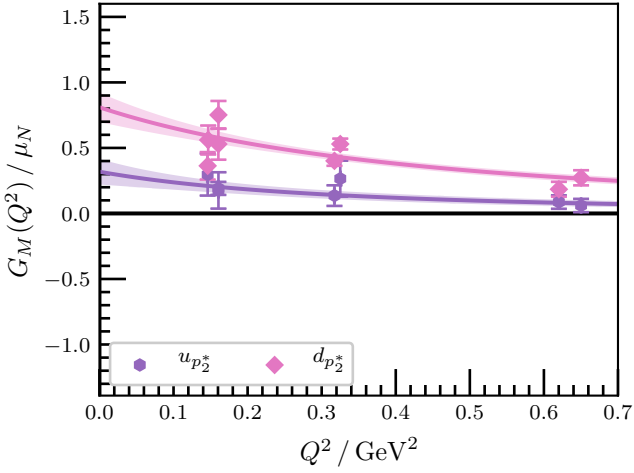}}
    \caption{\label{fig:2ndneg:GM:k3Q2qs}Quark-flavour contributions to
        \(\ffMagnetic\) for the second
        negative-parity excitation at \(m_{\pi} = \SI{411}{\mega\electronvolt}\).
        The curves are dipole fits to the individual quark sectors.
                                                }
\end{figure}

Once again the PEVA technique is crucial to extracting the correct form
factors. Hence, we focus only on the PEVA results.
Inspecting the \(Q^2\) dependence of these form factors, we find that the
contributions from both quark flavours agree well with a dipole ansatz. For
example, Fig.~\ref{fig:2ndneg:GM:k3Q2qs} shows the form factors at
\(m_{\pi} = \SI{411}{\mega\electronvolt}\). Here we have held the \(y\)-scale
fixed to match previous plots, for ease of comparison. The most notable feature
of these results is their small magnitude compared to the both the ground state
and the excitation considered in Section~\ref{sec:excitations:1stneg:GM}.

\subsubsection{Baryon magnetic form factors}

\begin{figure}[tbp]
    {\centering
        \includegraphics{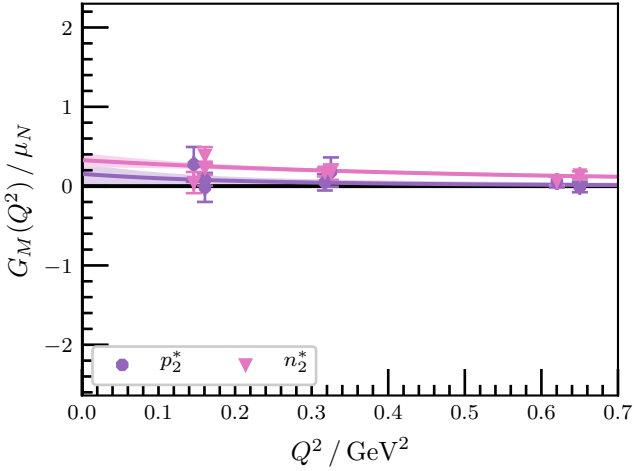}}
    \caption{\label{fig:2ndneg:GM:k3Q2nucleon}\(\ffMagnetic\) for the
        excited proton and neutron at \(m_{\pi} = \SI{411}{\mega\electronvolt}\).
        The curves correspond to combinations of the quark-sector dipole
        fits from Fig.~\ref{fig:2ndneg:GM:k3Q2qs}.                         }
\end{figure}

By taking linear combinations based on the multiplicity and charge of each
quark flavour, as described in Section~\ref{sec:excitations:1stneg:GE},
we can obtain the magnetic form factors for the excited proton
and neutron. Fig.~\ref{fig:2ndneg:GM:k3Q2nucleon} shows these
combinations for \(m_{\pi} = \SI{411}{\mega\electronvolt}\). The magnetic charge
radii\index{magnetic radius} obtained by combining the quark-sector dipoles are
consistent with the proton charge radii from \(\ffElectric\), although they
often have very large errors due to the very small values of the magnetic form
factors amplifying the effects of statistical fluctuations.

\subsubsection{Baryon magnetic moments}

\begin{figure}[tbp]
    {\centering
        \includegraphics{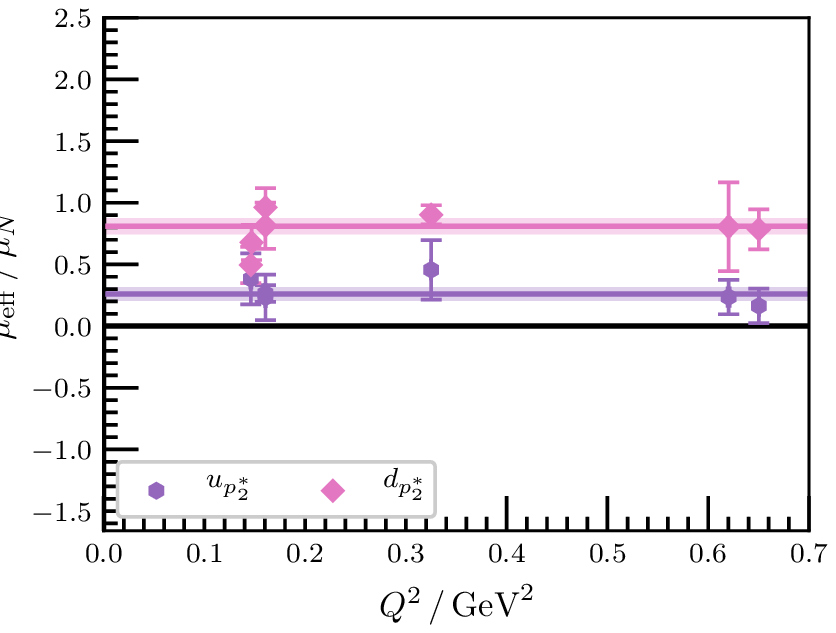}}
    \caption{\label{fig:2ndneg:mm:k3}\(\mmEff{}\) for individual quarks of unit
        charge in the second negative-parity excitation at
        \(m_{\pi} = \SI{411}{\mega\electronvolt}\). The shaded bands are
        constant fits to the effective magnetic moment,
        corresponding to magnetic moment contributions
        of \num{0.260 \pm 0.055} \(\mu_N\) for the doubly represented quark and
        \num{0.810 \pm 0.067} \(\mu_N\) for the singly represented quark.
        }
\end{figure}

Returning to the individual quark sector results with unit charge, and noting
that the \(Q^2\)
dependence for \(\ffElectric{}\) and \(\ffMagnetic{}\) is similar, we once
again take the ratio \(\mmEff{}(Q^2) \definedby \ffMagnetic{}/\ffElectric{}\).
We find that this ratio is approximately flat for all five pion masses.
For example, Fig.~\ref{fig:2ndneg:mm:k3} shows the \(Q^2\) dependence of the
ratio at \(m_{\pi} = \SI{411}{\mega\electronvolt}\). We can extract
the contributions to the magnetic moment\index{magnetic moment} from both quark flavours from constant
fits to this ratio.

\begin{figure}[tbp]
    {\centering
        \includegraphics{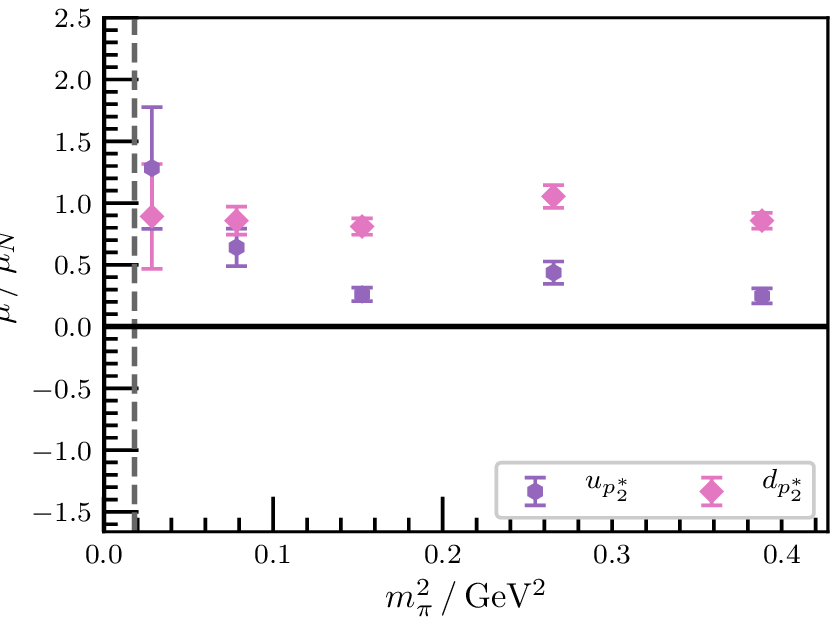}}
    \caption{\label{fig:2ndneg:mm:qs}Quark-mass dependence of contributions
        from individual unit-charge quarks to the magnetic moment of the second
        negative-parity excitation of the nucleon. The vertical
        dashed line corresponds to the physical pion mass.}
\end{figure}

Fig.~\ref{fig:2ndneg:mm:qs} shows the pion-mass dependence
of these extracted magnetic moment contributions. It is remarkable that both
quark flavours contribute with the same sign.

\begin{figure}[tbp]
    {\centering
        \includegraphics{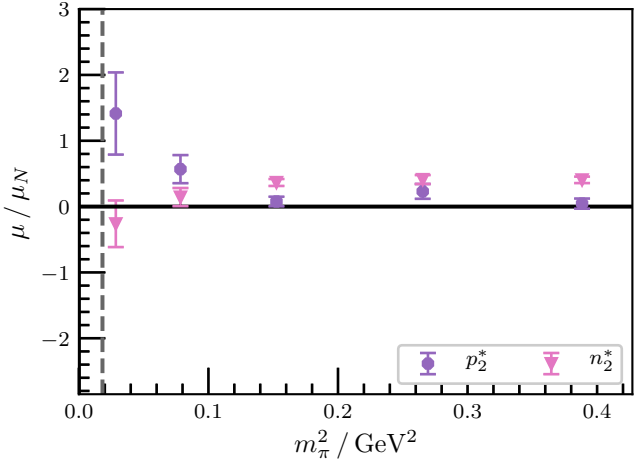}}
    \caption{\label{fig:2ndneg:mm:nucleon}Quark-mass dependence of
        the magnetic moment of the second negative-parity excitations of the proton and
        neutron. The dashed line corresponds to the physical pion mass. There is a clear
        change at the lightest two pion masses, which signals a significant
        shift in the structure of the states.}
\end{figure}

By taking the linear combinations discussed above, we can combine these
individual quark flavour results to get the predicted magnetic moments for the
second negative-parity excitations of the proton and neutron.
In Fig.~\ref{fig:2ndneg:mm:nucleon} we plot the dependence of these combinations
on the squared pion mass. For the heaviest three pion masses, the effective
magnetic moments\index{magnetic moment} show little pion mass dependence and have tight error bars.
The magnetic moment of the proton excitation sits very close to zero, and the
magnetic moment of the neutron excitation has a small but nonzero positive value.
For the lightest two masses, the ordering
of the two states flips, with the proton excitation taking on a more significant
magnetic moment, and the neutron excitation dropping to be consistent with or
below zero.

In summary, the PEVA technique is crucial for extracting both the electric
and magnetic form factors of the second negative-parity excitations of the
proton and neutron.
In Table~\ref{tab:formfactors:state2},
we present the charge radii, magnetic radii, and magnetic moments of the second
negative-parity excitation of the proton and neutron. We see
that this state has similar radii to the ground state, but notably different
magnetic moments. We find that at the heavier quark masses, these magnetic
moments agree well with constituent quark model predictions, as discussed
below.

\begin{table}[hbt]
    \caption{\label{tab:formfactors:state2}Radii and magnetic
    moments of the second negative-parity excitation of the proton and neutron.
    Radii are obtained from from combinations of quark-sector dipole fits and magnetic moments are obtained
    from quark-sector ratios of \(\ffMagnetic\) to \(\ffElectric{}\).
    At the lightest two pion masses, the form factor data was
    insufficient to properly constrain a dipole fit so we do not
    report magnetic radii at these masses.
    }
    \sisetup{%
    table-number-alignment = center,
    table-figures-integer = 1,
    table-figures-decimal = 0,
    table-figures-uncertainty = 0,
}
\begin{tabular*}{\linewidth}{%
@{\extracolsep{\fill}}
        S[table-figures-integer = 1, table-figures-decimal = 4, table-figures-uncertainty = 3]
        S[table-figures-integer = 1, table-figures-decimal = 3, table-figures-uncertainty = 2]
        S[table-figures-integer = 1, table-figures-decimal = 2, table-figures-uncertainty = 4]
        S[table-figures-integer = 1, table-figures-decimal = 2, table-figures-uncertainty = 2]
}
    \toprule{} \vspace{-9pt}\\
    {\(m_{\pi}^2 \, / \, \si{\giga\electronvolt^2}\)} & {\(\rsqElectricSprsb{} \, / \, \si{\femto\meter\squared}\)} & {\(\rsqMagneticSprsb{} \, / \, \mmSprsb{}\! \, / \, \si{\femto\meter\squared}\)} & {\(\mmSprsb{}\! \, / \, \mu_N\)} \\
    \colrule{} \vspace{-9pt}\\
    0.3884 \pm 0.0113 & 0.363 \pm 0.035 & 0.56 \pm 2.13 & 0.05 \pm 0.08 \\
    0.2654 \pm 0.0081 & 0.399 \pm 0.029 & 0.55 \pm 0.70 & 0.23 \pm 0.11 \\
    0.1525 \pm 0.0043 & 0.445 \pm 0.062 & 1.09 \pm 0.70 & 0.08 \pm 0.07 \\
    0.0784 \pm 0.0025 & 0.599 \pm 0.077 & {\textemdash} & 0.57 \pm 0.21 \\
    0.0285 \pm 0.0012 & 0.428 \pm 0.086 & {\textemdash} & 1.41 \pm 0.62 \\
    \botrule{}
\end{tabular*}
    \sisetup{%
    table-number-alignment = center,
    table-figures-integer = 1,
    table-figures-decimal = 0,
    table-figures-uncertainty = 0,
}
\begin{tabular*}{\linewidth}{%
@{\extracolsep{\fill}}
        S[table-figures-integer = 1, table-figures-decimal = 4, table-figures-uncertainty = 3]
        S[table-sign-mantissa,table-figures-integer = 2, table-figures-decimal = 3, table-figures-uncertainty = 2]
        S[table-figures-integer = 1, table-figures-decimal = 2, table-figures-uncertainty = 2]
        S[table-sign-mantissa,table-figures-integer = 2, table-figures-decimal = 2, table-figures-uncertainty = 2]
}
    \toprule{} \vspace{-9pt}\\
    {\(m_{\pi}^2 \, / \, \si{\giga\electronvolt^2}\)} & {\(\rsqElectricSnesb{} \, / \, \si{\femto\meter\squared}\)} & {\(\rsqMagneticSnesb{} \, / \, \mmSnesb{}\! \, / \, \si{\femto\meter\squared}\)} & {\(\mmSnesb{}\! \, / \, \mu_N\)} \\
    \colrule{} \vspace{-9pt}\\
    0.3884 \pm 0.0113 & 0.014 \pm 0.017 & 0.30 \pm 0.18 & 0.41 \pm 0.05 \\
    0.2654 \pm 0.0081 & 0.003 \pm 0.014 & 0.65 \pm 0.30 & 0.41 \pm 0.07 \\
    0.1525 \pm 0.0043 & 0.033 \pm 0.030 & 0.41 \pm 0.34 & 0.37 \pm 0.05 \\
    0.0784 \pm 0.0025 & -0.076 \pm 0.047 & {\textemdash} & 0.14 \pm 0.14 \\
    0.0285 \pm 0.0012 & 0.036 \pm 0.068 & {\textemdash} & -0.26 \pm 0.35 \\
    \botrule{}
\end{tabular*}
\end{table}

\subsection{Model comparison\label{sec:excitations:model}}

\begin{figure*}[tb]
    {\centering
        \includegraphics{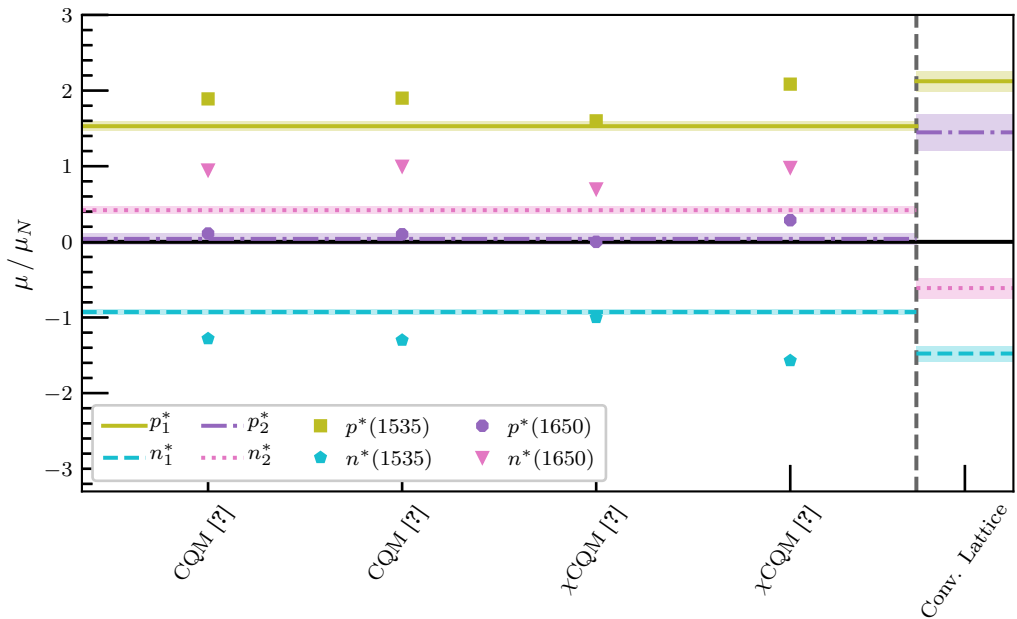}}
    \caption{\label{fig:excitations:model:k1}Comparison between lattice
    calculations of the magnetic moments of two negative-parity nucleon
    excitations at \(m_{\pi} = \SI{702}{\mega\electronvolt}\)
    and quark model predictions for the \(N^*(1535)\)    \index{N*1535@\(N^*(1535)\)} and \(N^*(1650)\) resonances    \index{N*1650@\(N^*(1650)\)}. The shaded bands on the left-hand side of the
    plot indicate the
    magnetic moments calculated via the PEVA technique,
    and symbols denote the quark model predictions.
    Lattice calculations of the magnetic moments using a
    conventional variational analysis are plotted to the right of the vertical
    dashed line.
    }
\end{figure*}

In Section~\ref{sec:excitations:introduction}, we introduced the two low-lying negative-parity
excitations of the proton and neutron observed on the lattice by the CSSM and the HSC
collaborations.
We wish to examine the extent to which these states, formed in relativistic quantum field theory,
have properties that are captured in simple quark models.  As these states have good overlap with
local three-quark operators, we anticipate these states may resemble the quark-model states
postulated to describe these resonances \cite{Chiang:2002ah,Liu:2005wg,Sharma:2013rka}.

Already we have seen some deviations from quark model expectations for electric charge radii.
There, larger radii for the $N^*$ states were anticipated due to centripetal repulsion in the radial
Schr\"odinger equation for three-quark states with orbital angular momentum $\ell = 1$.  However,
similar charge radii point to five-quark meson-baryon components in the $N^*$ structure where the
anti-quark provides negative parity and all quarks can sit in relative $s$ waves.

Here, we focus on the magnetic moments\index{magnetic moment} of these two negative-parity states, as calculated
in sections~\ref{sec:excitations:1stneg:GM} and~\ref{sec:excitations:2ndneg:GM}.  It will be
interesting to learn if there is a nontrivial role for meson-baryon Fock-space components here as well.

Our focus on the three heavier quark masses considered is beneficial in comparing with quark models
as it is this regime where constituent quark phenomenology is expected to be manifest
\cite{Leinweber:1990dv,Leinweber:1991vc}.

We consider two constituent quark model\index{constituent quark model} (QM) predictions of
the magnetic moments from Refs.~\cite{Chiang:2002ah,Liu:2005wg}, and two chiral
constituent quark model (\(\chi\)QM) calculations which take the quark model calculations
and include effects from the pion cloud~\cite{Liu:2005wg,Sharma:2013rka}, thus incorporating
meson-baryon Fock-space components.

In Fig.~\ref{fig:excitations:model:k1}, we compare our magnetic moments extracted
at \(m_{\pi} = \SI{702}{\mega\electronvolt}\) with these quark model predictions,
which are calculated at the physical point. We can see that qualitatively, the
results for the first negative-parity lattice excitation match up with the quark
model \(N^*(1535)\)\index{N*1535@\(N^*(1535)\)}, and the second negative-parity
lattice excitation with the quark model \(N^*(1650)\)\index{N*1650@\(N^*(1650)\)}.
In fact, despite being at significantly different pion masses, the results are
quantitatively very similar, with the lattice results sitting within the scatter
of the model predictions for all states save the second negative-parity nucleon
excitation, which sits slightly below all of the model predictions.

For comparison, we also plot lattice results produced using the conventional
variational analysis. For these results, \(\mmEff{}(Q^2)\) varies significantly for
different kinematics, so rather than taking a constant fit across kinematics,
we present only the result from the lowest-momentum (\(\vect{p} = (0,0,0)\),
\(\vect{p}' = (1,0,0)\))
kinematics, which we expect to have the smallest opposite-parity contaminations.
We see that the conventional results are significantly different to the PEVA
results. In particular, the conventional extraction of the second
negative-parity excitation is completely different to both the PEVA result and the
quark model results. This once again demonstrates how critical the PEVA
technique is to obtaining correct results.

\begin{figure*}[tb]
    {\centering
        \includegraphics{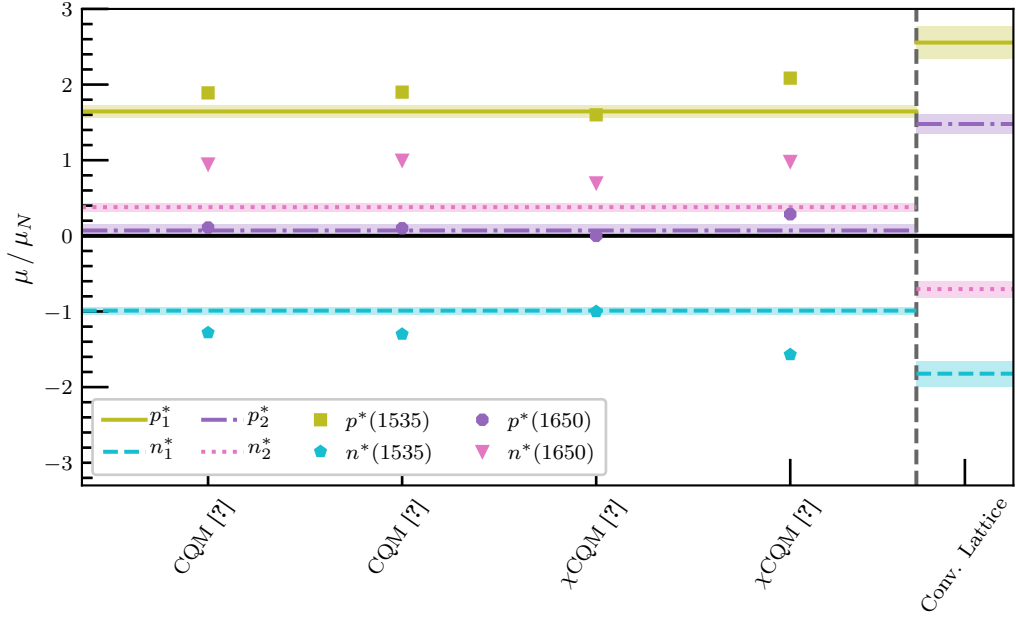}}
    \caption{\label{fig:excitations:model:k3}Comparison between magnetic moments
    from lattice calculations at \(m_{\pi} = \SI{411}{\mega\electronvolt}\)
    and quark model predictions for the \(N^*(1535)\)    \index{N*1535@\(N^*(1535)\)} and \(N^*(1650)\)\index{N*1650@\(N^*(1650)\)}.
    The shaded bands indicate the PEVA calculations on the left, and the
    conventional analysis on the right. The markers
    show the quark model predictions.
    }
\end{figure*}

This trend continues for \(m_{\pi} = \SI{570}{\mega\electronvolt}\), and
\(m_{\pi} = \SI{411}{\mega\electronvolt}\), the latter shown in
Fig.~\ref{fig:excitations:model:k3}. Since the pion-mass dependence of the
magnetic moments between these three masses is quite small, the quantitative
agreement remains good.

\begin{figure*}[tb]
    {\centering
        \includegraphics{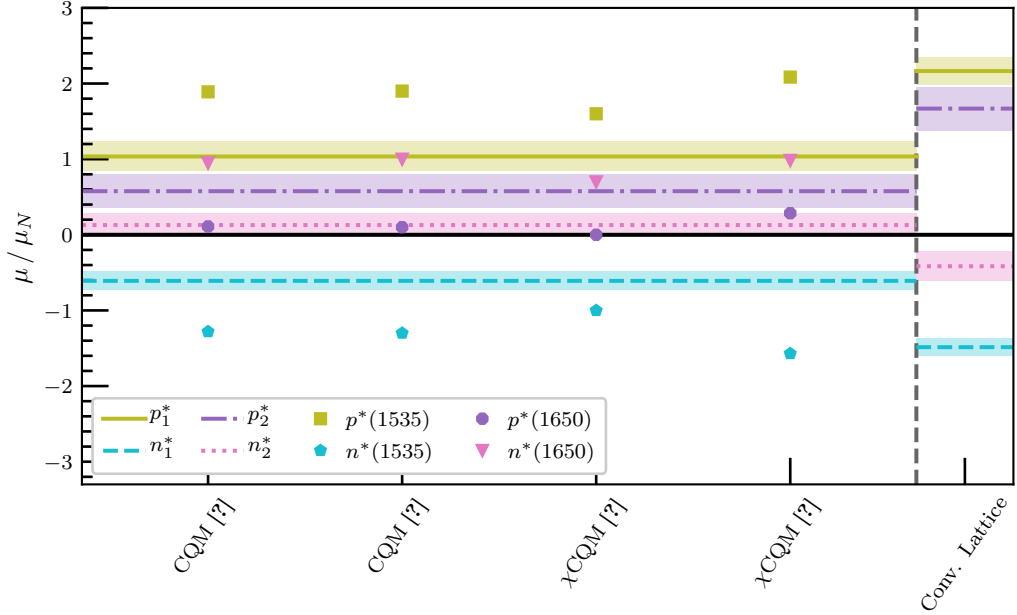}}
    \caption{\label{fig:excitations:model:k4}Comparison between magnetic moments from lattice
      calculations at \(m_{\pi} = \SI{296}{\mega\electronvolt}\) and quark model predictions for
      the \(N^*(1535)\)\index{N*1535@\(N^*(1535)\)} and \(N^*(1650)\)\index{N*1650@\(N^*(1650)\)}.
      The shaded bands indicate the PEVA calculations on the left, and the conventional analysis on
      the right. The markers show the quark model predictions.  }
\end{figure*}

For completeness, we also present a comparison for the second lightest mass.
Fig.~\ref{fig:excitations:model:k4} shows that the lattice results depart from the model
predictions at \(m_{\pi} = \SI{296}{\mega\electronvolt}\).  Again, an analysis of possible
scattering-state contributions to the correlation functions is required to disentangle interesting
meson-cloud effects from scattering-state contaminations as discussed in
  Sec.~\ref{subsec:multi-part}.
Fig.~\ref{fig:excitations:model:k4} also shows that there is still a significant
disagreement between the conventional and PEVA results at
\(m_{\pi} = \SI{296}{\mega\electronvolt}\). It remains clear that the PEVA
technique is playing an important role in addressing opposite parity
contaminations.

The results presented in this section provide new insight into the structure of the negative-parity
nucleon excitations observed in lattice quantum field theory.  At the heavier quark masses
considered, the two negative-parity excitations agree well with quark-model descriptions for the
\(N^*(1535)\)\index{N*1535@\(N^*(1535)\)} and \(N^*(1650)\)\index{N*1650@\(N^*(1650)\)}.  Coupled
with the charge radii from Section~\ref{sec:excitations:1stneg:GE}, which indicate
the importance of meson-nucleon components in the wave function, and the observed single-particle
dispersion relations seen in Ref.~\cite{Menadue:2013kfi}, the results indicate that these states
are similar in structure to the ground state nucleon which can also be modelled as a three-quark
state dressed by a meson cloud.

\bigskip

\section{Positive parity excitation\label{sec:excitations:1stpos}}

We now move to the positive parity sector, studying the first localised positive-parity
excitation of the nucleon observed on the lattice. This state sits at an
effective mass\index{effective energy} of approximately \SI{2}{\giga\electronvolt} for all five pion
masses, well above \SI{1.43(2)}{\giga\electronvolt}, the mass of the Roper
resonance observed in nature~\cite{Patrignani:2016xqp}. This has long been a
puzzle for the particle physics community, but recent HEFT results indicate
that the Roper resonance is dynamically generated from meson-baryon
scattering states~\cite{Liu:2016uzk}, and hence the lattice spectrum in this
energy region has poor overlap with local three-quark operators. This means
that the lattice state studied here is likely associated with the \(N^*(1710)\),
\(N^*(1880)\), and/or \(N^*(2100)\) resonances.

\subsection{Electric form factor}

\subsubsection{Quark-flavour contributions}

\begin{figure}[tbp]
    {\centering
        \includegraphics{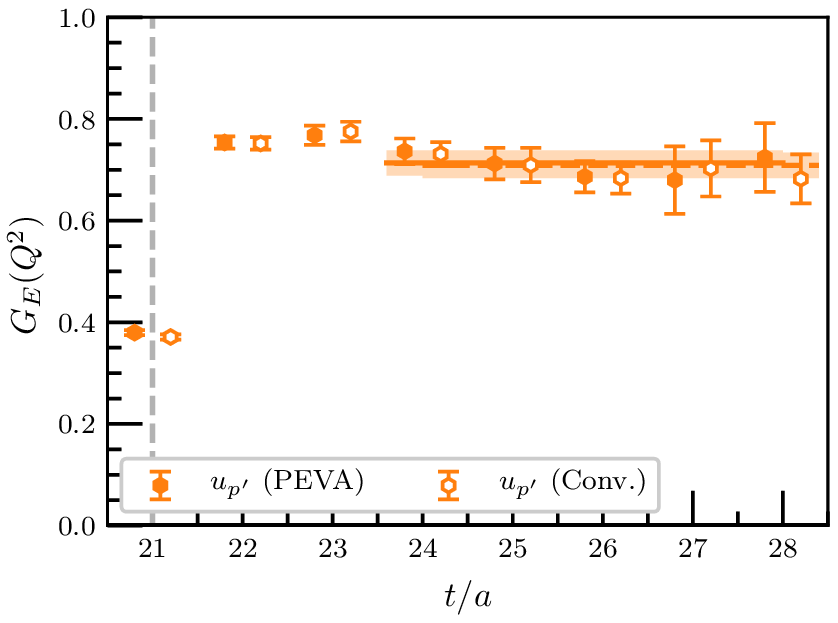}}
    \caption{\label{fig:1stpos:GEqs1:k1p000pp100}Contributions to \(\ffElectric\) from the doubly
      represented quark flavour for the first positive-parity nucleon excitation at \(m_{\pi} =
      \SI{702}{\mega\electronvolt}\) with the lowest-momentum kinematics, providing \(Q^2 =
      \SI{0.1425 \pm 0.0041}{\giga\electronvolt^2}\). We plot the conventional analysis with open
      markers and dashed fit lines and the PEVA technique with filled markers and solid fit lines.
      Both fits are from time slice \num{24}--\num{28}.  }
\end{figure}

\begin{figure}[tbp]
    {\centering
        \includegraphics{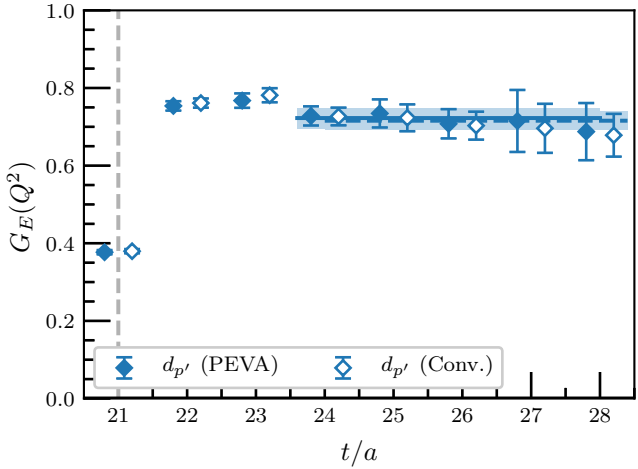}}
    \caption{\label{fig:1stpos:GEqs2:k1p000pp100}Contributions to \(\ffElectric\)
        from the singly represented quark flavour
        for the first positive-parity nucleon excitation.
        The pion mass, kinematics and plotting convention are the same as in
        Fig.~\ref{fig:1stpos:GEqs1:k1p000pp100} above.
        Both fits are from time slice \num{24}--\num{28}.
        }
\end{figure}

We plot the dependence of \(\ffElectric\) on the Euclidean sink time at
\(m_{\pi} = \SI{702}{\mega\electronvolt}\) in
Figs.~\ref{fig:1stpos:GEqs1:k1p000pp100} and~\ref{fig:1stpos:GEqs2:k1p000pp100}.
The form factor values extracted from the PEVA and conventional analyses for
each sink time look very similar, and this is reflected in the fits.
The conventional and PEVA extractions both have clear plateaus over the same
range of sink times, and these plateaus have consistent values.

This trend continues for lighter pion masses:
the PEVA and conventional analyses have the same fit ranges and consistent fit
values. This is also true for all kinematics for which we are able to find
acceptable plateaus. This suggests there are no significant effects from
opposite-parity contaminations on \(\ffElectric{}\) for this state, at
least at this level of statistics.

\begin{figure}[tb]
    {\centering
        \includegraphics{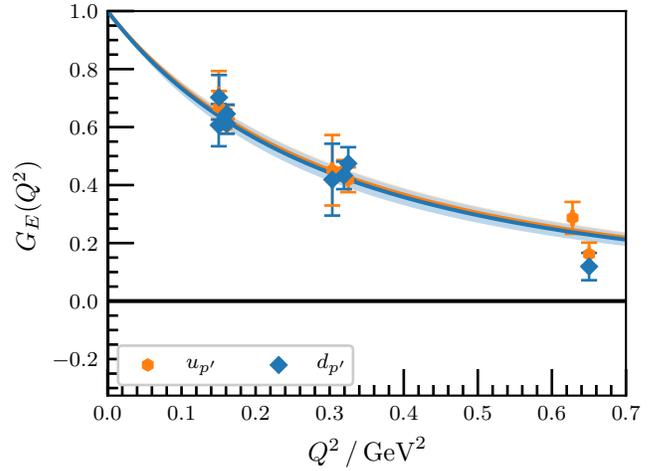}}
    \caption{\label{fig:1stpos:GE:k3Q2qs}Quark-flavour contributions to
        \(\ffElectric\) for the first positive-parity excitation at
        \(m_{\pi} = \SI{411}{\mega\electronvolt}\). The
        curves are dipole fits to the form factor, with lines
        indicating the central values. The fits correspond to RMS charge
        radii of \SI{0.871 \pm 0.036}{\femto\meter} for the doubly represented quark
        flavour (\(\indSdblp\)) and \SI{0.885 \pm 0.044}{\femto\meter} for the singly
        represented quark flavour (\(\indSsingp\)).
        }
\end{figure}

Focusing on the PEVA results, in Fig.~\ref{fig:1stpos:GE:k3Q2qs}, we plot the
\(Q^2\) dependence of the electric form factor for the two valence quark flavours
at \(m_{\pi} = \SI{411}{\mega\electronvolt}\). We see that the two quark
flavours make very similar contributions to the electric form factor and agree well
with a dipole fit corresponding to charge radii\index{charge radius} of \SI{0.88(4)}{\femto\meter}
for the doubly represented quark flavour and \SI{0.89(5)}{\femto\meter} for the
singly represented quark flavour. This is significantly larger than the
ground-state nucleon. Similar behaviour is seen for the other four
masses.

\begin{figure}[tbp]
    {\centering
        \includegraphics{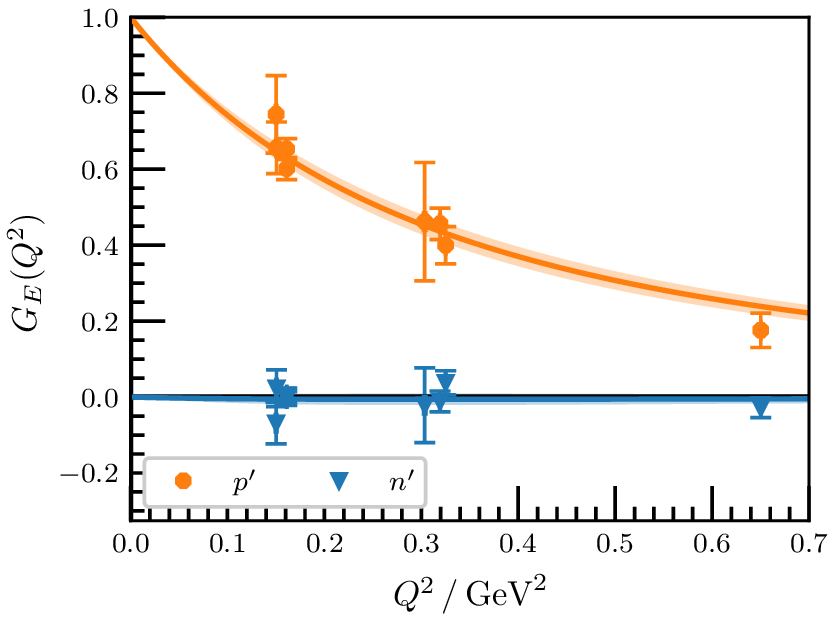}}
    \caption{\label{fig:1stpos:GE:k3Q2nucleon}\(\ffElectric\) for the
        first positive-parity excitations of the proton and neutron
        at \(m_{\pi} = \SI{411}{\mega\electronvolt}\).
        The curves correspond to combinations of the quark-sector dipole
        fits from Fig.~\ref{fig:1stpos:GE:k3Q2qs}, giving a squared charge radius of
        \SI{0.75 \pm 0.07}{\femto\meter\squared} for the proton
        and \SI{0.016 \pm 0.039}{\femto\meter\squared} for the neutron.
        }
\end{figure}

\subsubsection{Baryon electric form factors}

As above, we take linear combinations of the individual quark flavour contributions,
including the charges of the quark flavours and their multiplicity, to get the
electric form factors for the first positive-parity excitations of the proton
and neutron.
In Fig.~\ref{fig:1stpos:GE:k3Q2nucleon}, we plot these combinations at
\(m_{\pi} = \SI{411}{\mega\electronvolt}\). At this and the other four masses,
we find that the electric form factor for the neutron excitation is
approximately zero.

\begin{figure}[tbp]
    {\centering
        \includegraphics{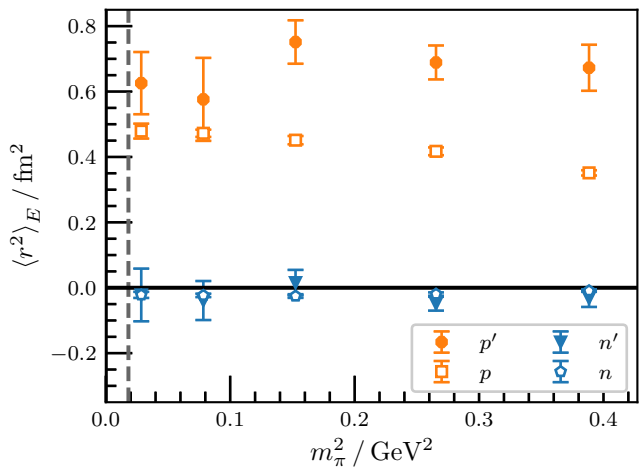}}
    \caption{\label{fig:1stpos:GE:proton}Quark-mass dependence of charge radii
    from combinations of quark-sector dipole fits to \(\ffElectric\) for the first positive-parity
    excitation of the proton.}
\end{figure}

In Fig.~\ref{fig:1stpos:GE:proton}, we plot the pion-mass dependence of charge radii
extracted from combinations of the quark-sector dipole fits to the electric form factor.  For the
heaviest three masses, squared charge radii\index{charge radius} for the proton range from \SI{0.67
  \pm 0.07}{\femto\meter\squared} to \SI{0.75 \pm 0.07}{\femto\meter\squared}, increasing with
decreasing pion mass. These radii are all significantly larger than the charge radius of the
ground-state proton at the corresponding mass.
Thus the second positive-parity excitation is
significantly larger than the ground state proton at these pion masses, in line with
earlier observations that this positive parity excitation has a wave function consistent with a
three-quark radial excitation with one node \cite{Roberts:2013ipa,Roberts:2013oea}.
At the lightest two masses, the statistical errors become large.  However the trend for this
positive parity excitation to be larger than the ground state is manifest.

\subsection{Magnetic form factor}

\subsubsection{Quark-flavour contributions}

\begin{figure}[bp]
    {\centering
        \includegraphics{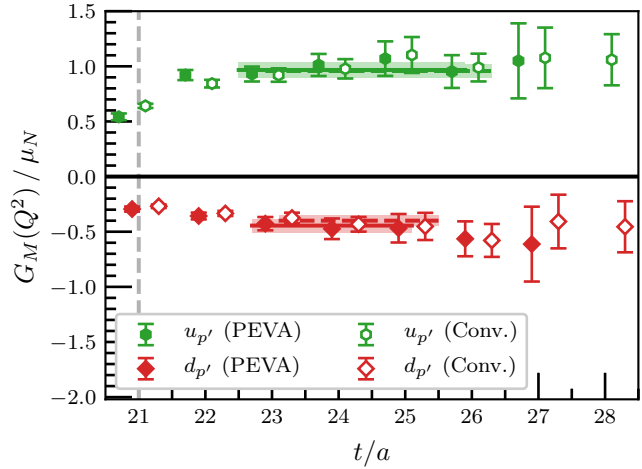}}
    \caption{\label{fig:1stpos:GM:k1p000pp100}Quark-flavour contributions to
        the magnetic form factor for the first
        positive-parity excitation of the nucleon at
        \(m_{\pi} = \SI{702}{\mega\electronvolt}\) for the lowest-momentum
        kinematics, providing \(Q^2 = \SI{0.1425 \pm 0.0041}{\giga\electronvolt^2}\).         Results are for single quarks of unit charge. All four fits start from
        time slice \num{23} and have consistent values.
        }
\end{figure}

\begin{figure}[tb]
    {\centering
        \includegraphics{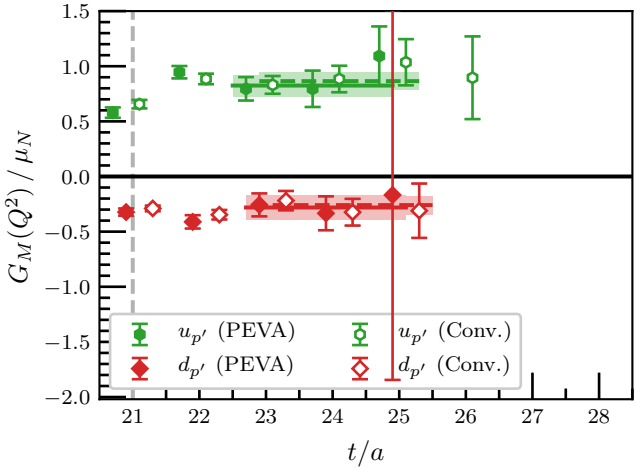}}
    \caption{\label{fig:1stpos:GM:k3p000pp100}Quark-flavour contributions to
        \(\ffMagnetic\) for the first
        positive-parity excitation of the nucleon at
        \(m_{\pi} = \SI{411}{\mega\electronvolt}\) for the lowest-momentum
        kinematics, providing \(Q^2 = \SI{0.146}{\giga\electronvolt^2}\).
        }
\end{figure}

Having investigated the electric form factor for this state, we now consider
the magnetic form factor. In Fig.~\ref{fig:1stpos:GM:k1p000pp100} we plot the
Euclidean sink-time dependence of the extracted form factors at
\(m_{\pi} = \SI{702}{\mega\electronvolt}\), with the lowest-momentum kinematics.
We see that the form factors and plateaus for both analyses are very similar,
and there is no evidence for opposite-parity contamination of this state. We see
similar results for the other masses and kinematics, with no clear differences
between the conventional and PEVA plateaus. For example,
Fig.~\ref{fig:1stpos:GM:k3p000pp100}
shows this behaviour at \(m_{\pi} = \SI{411}{\mega\electronvolt}\) with the
same lowest-momentum kinematics. This suggests that, like
\(\ffElectric\), \(\ffMagnetic\) for the first positive-parity excitation is
not affected by opposite parity excitations, at least at this level of
statistics.

\begin{figure}[tbp]
    {\centering
        \includegraphics{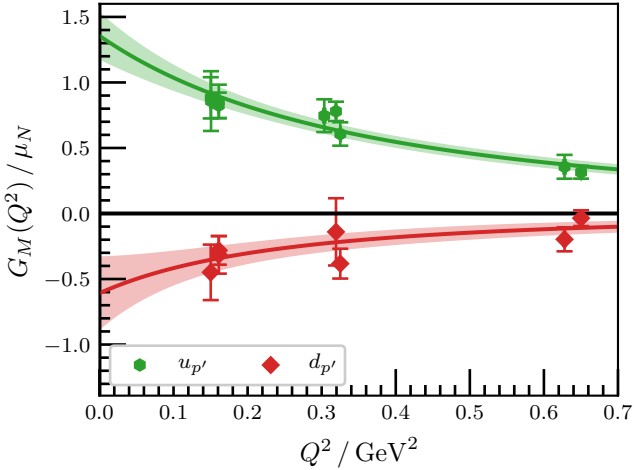}}
    \caption{\label{fig:1stpos:GM:k3Q2qs}Quark-flavour contributions to
        \(\ffMagnetic\) for the first
        positive-parity excitation at \(m_{\pi} = \SI{411}{\mega\electronvolt}\).
        The curves are dipole fits to the form factor,
        corresponding to squared magnetic radii of \SI{0.67 \pm 0.14}{\femto\meter\squared}
        for \(\indSdblp\) and
        \SI{0.97 \pm 0.59}{\femto\meter\squared} for \(\indSsingp\).
        }
\end{figure}

Focusing on the PEVA results, we plot the \(Q^2\) dependence of the plateau
fits for the two valence quark flavours at \(m_{\pi} = \SI{411}{\mega\electronvolt}\)
in Fig.~\ref{fig:1stpos:GM:k3Q2qs}. We see that both quark flavours agree well
with a dipole ansatz. This is also true for the two heavier pion masses, and
the two lighter masses are also consistent, though they
are too noisy to significantly constrain the fit.

\begin{figure}[tbp]
    {\centering
        \includegraphics{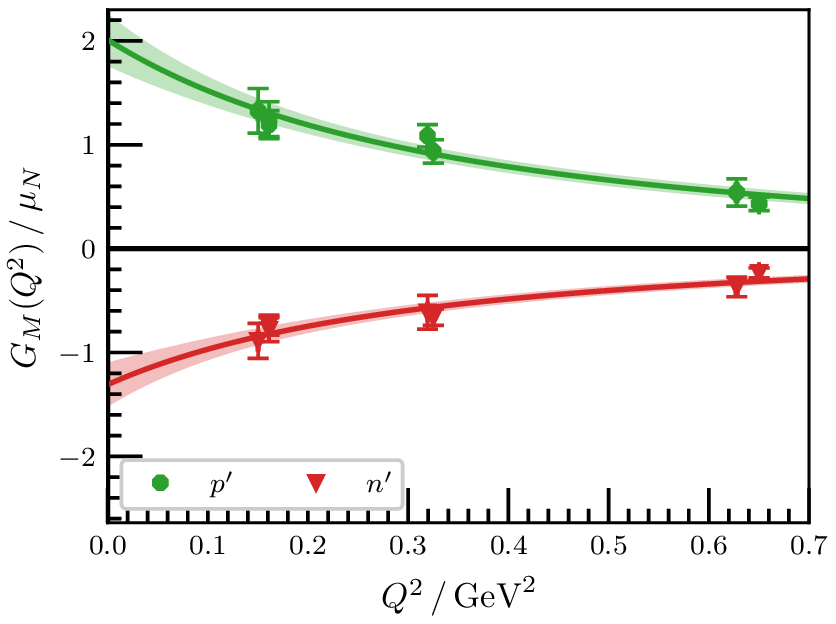}}
    \caption{\label{fig:1stpos:GM:k3Q2nucleon}\(\ffMagnetic\) for the
        first positive-parity excitations of the proton and neutron
        at \(m_{\pi} = \SI{411}{\mega\electronvolt}\).                         The curves correspond to combinations of the quark-sector dipole fits from Fig.~\ref{fig:1stpos:GM:k3Q2qs},
        giving a squared magnetic radius of \SI{0.70 \pm 0.13}{\femto\meter\squared} for the
        proton excitation and \SI{0.77 \pm 0.21}{\femto\meter\squared} for the neutron excitation.
        }
\end{figure}

\subsubsection{Baryon magnetic form factors}

\begin{figure}[tbp]
    {\centering
        \includegraphics{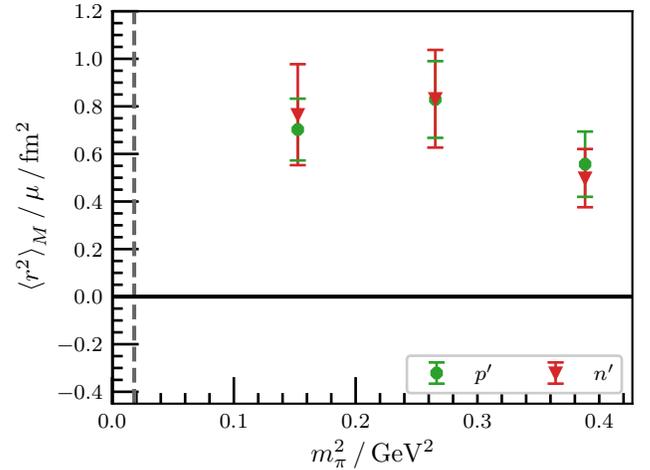}}
    \caption{\label{fig:1stpos:GM:radius}Quark-mass dependence of squared magnetic
        radii of the first positive-parity excitation of the proton and neutron
        from quark-sector dipole fits to \(\ffMagnetic{}\).
    At the lightest two pion masses, the form factor data was
    insufficient to properly constrain a dipole fit so we do not
    report magnetic radii at these masses.
        }
\end{figure}

Combinations of the quark flavour contributions are taken to form the proton and neutron
excitations.  In Fig.~\ref{fig:1stpos:GM:k3Q2nucleon}, we plot these combinations at \(m_{\pi} =
\SI{411}{\mega\electronvolt}\). By combining the dipole fits to the quark-sector results in the
same way, we obtain magnetic radii\index{magnetic radius} that are consistent with the
corresponding excited proton charge radius.  This can be seen in Fig.~\ref{fig:1stpos:GM:radius},
in which we plot the pion-mass dependence of the squared magnetic radii obtained from the linear
combinations of the dipole fits to the form factors. These plots show fairly consistent results for
the heavier three masses, with some pion-mass dependence. We omit results for the lightest two
masses because the form factor data is insufficient to constrain dipole fits at these masses.

\subsubsection{Baryon magnetic moments}

\begin{figure}[tbp]
    {\centering
        \includegraphics{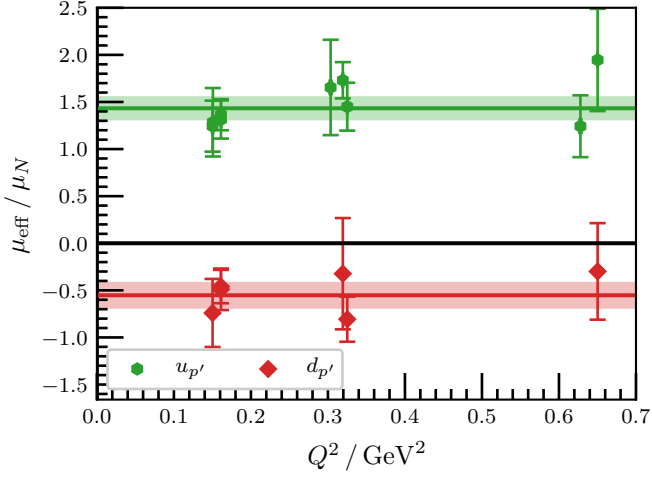}}
    \caption{\label{fig:1stpos:mm:k3}\(\mmEff{}\) for individual quarks of unit
        charge in the first
        positive-parity excitation at \(m_{\pi} = \SI{411}{\mega\electronvolt}\).
        The shaded bands are
        constant fits to the effective magnetic moments which provide
        magnetic moment contributions
        of \num{1.43 \pm 0.13} \(\mu_N\) for the doubly represented quark and
        \num{-0.55 \pm 0.14} \(\mu_N\) for the singly represented quark.
        }
\end{figure}

Returning to the individual quark sector results and noting that once again the
electric and magnetic form factors have a similar \(Q^2\) dependence, we take
the ratio \(\mmEff{}(Q^2) \definedby \ffMagnetic{} / \ffElectric{}\). In
Fig.~\ref{fig:1stpos:mm:k3}, we plot this ratio as a function of \(Q^2\) for
\(m_{\pi} = \SI{411}{\mega\electronvolt}\). We
find that the ratio is once again very flat in \(Q^2\), supporting our
hypothesis that the form factors
have the same \(Q^2\) scaling in this region, and the validity of \(\mmEff{}\)
as an estimate of the magnetic moment\index{magnetic moment}.

\begin{figure}[tbp]
    {\centering
        \includegraphics{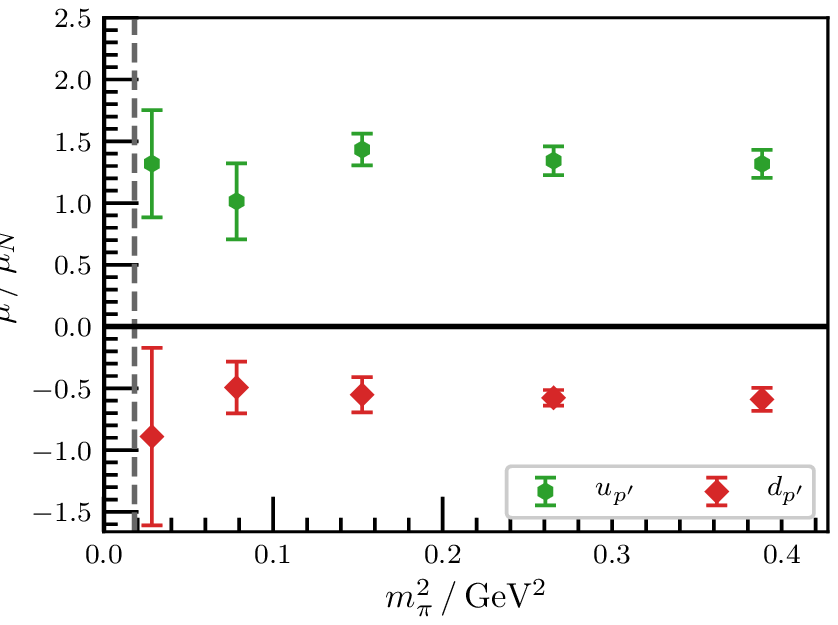}}
    \caption{\label{fig:1stpos:mm:qs}Quark-mass dependence of contributions
        from individual unit-charge quarks to the magnetic moment of the
        first positive-parity excitation of the nucleon. The vertical
        dashed line corresponds to the physical pion mass.}
\end{figure}

\begin{figure}[tbp]
    {\centering
        \includegraphics{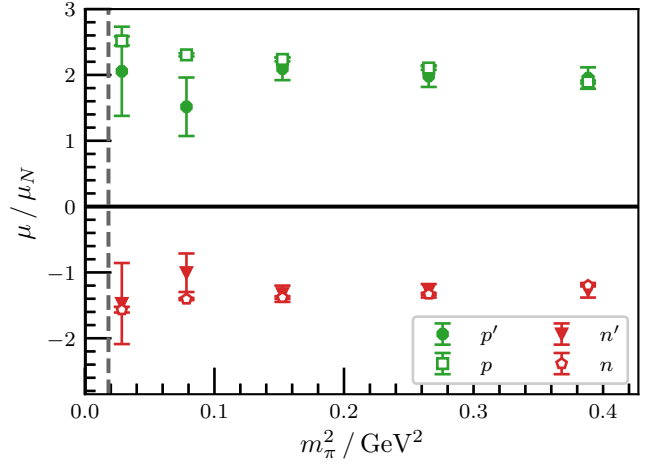}}
    \caption{\label{fig:1stpos:mm:nucleon}Quark-mass dependence of the magnetic
        moments of the first positive-parity excitations of the proton and
        neutron.
        The vertical dashed line corresponds to the physical pion mass, and
        the ground state magnetic moments have been plotted with open markers.}
\end{figure}

In Fig.~\ref{fig:1stpos:mm:qs}, we plot the pion-mass dependence of
\(\mmEff{}\) for individual quarks of unit charge.
We can once again take combinations of the individual quark-flavour
contributions to get the excited proton and neutron magnetic moments.
In Fig.~\ref{fig:1stpos:mm:nucleon}, we plot the pion-mass dependence of these
combinations.

We see that the excited-state magnetic moments agree well with the ground-state magnetic
moments\index{magnetic moment}, particularly at the heaviest quark masses, where the agreement is
impressive.  These results are in accord with a simple $2S$ constituent-quark-model\index{constituent
  quark model} state.

In summary, we have shown that the first positive-parity excitation of the nucleon has no obvious
opposite-parity contaminations. However, variational analysis techniques in general have provided
good access to this state at several pion masses. This has allowed us to ascertain for the first
time that these states have a larger radius\index{charge radius} than the ground-state nucleon, but
have very similar magnetic moments\index{magnetic moment}. This is consistent with these states
being dominated by a radial $2S$ excitation of the ground-state nucleon as seen in
Refs.~\cite{Roberts:2013ipa,Roberts:2013oea}.  Table~\ref{tab:formfactors:state3}, collects our
charge radii, magnetic radii, and magnetic moments for this positive-parity excitation of the
proton and neutron.

\begin{table}[hbt]
    \caption{\label{tab:formfactors:state3}Radii and magnetic
    moments of the positive-parity excitation of the proton and neutron.
    Radii are obtained from from combinations of quark-sector dipole fits and magnetic moments are obtained
    from quark-sector ratios of \(\ffMagnetic\) to \(\ffElectric{}\).
    At the lightest two pion masses, the form factor data was
    insufficient to properly constrain a dipole fit so we do not
    report magnetic radii at these masses.
    }
    \sisetup{%
    table-number-alignment = center,
    table-figures-integer = 1,
    table-figures-decimal = 0,
    table-figures-uncertainty = 0,
}
\begin{tabular*}{\linewidth}{%
@{\extracolsep{\fill}}
        S[table-figures-integer = 1, table-figures-decimal = 4, table-figures-uncertainty = 3]
        S[table-figures-integer = 1, table-figures-decimal = 3, table-figures-uncertainty = 3]
        S[table-figures-integer = 1, table-figures-decimal = 2, table-figures-uncertainty = 2]
        S[table-figures-integer = 1, table-figures-decimal = 2, table-figures-uncertainty = 2]
}
    \toprule{} \vspace{-9pt}\\
    {\(m_{\pi}^2 \, / \, \si{\giga\electronvolt^2}\)} & {\(\rsqElectricSprp{} \, / \, \si{\femto\meter\squared}\)} & {\(\rsqMagneticSprp{} \, / \, \mmSprp{}\! \, / \, \si{\femto\meter\squared}\)} & {\(\mmSprp{}\! \, / \, \mu_N\)} \\
    \colrule{} \vspace{-9pt}\\
    0.3884 \pm 0.0113 & 0.673 \pm 0.073 & 0.56 \pm 0.14 & 1.95 \pm 0.17 \\
    0.2654 \pm 0.0081 & 0.689 \pm 0.056 & 0.83 \pm 0.16 & 1.98 \pm 0.17 \\
    0.1525 \pm 0.0043 & 0.751 \pm 0.069 & 0.70 \pm 0.13 & 2.09 \pm 0.18 \\
    0.0784 \pm 0.0025 & 0.576 \pm 0.128 & {\textemdash} & 1.52 \pm 0.44 \\
    0.0285 \pm 0.0012 & 0.626 \pm 0.097 & {\textemdash} & 2.06 \pm 0.68 \\
    \botrule{}
\end{tabular*}
    \sisetup{%
    table-number-alignment = center,
    table-figures-integer = 1,
    table-figures-decimal = 0,
    table-figures-uncertainty = 0,
}
\begin{tabular*}{\linewidth}{%
@{\extracolsep{\fill}}
        S[table-figures-integer = 1, table-figures-decimal = 4, table-figures-uncertainty = 3]
        S[table-sign-mantissa,table-figures-integer = 2, table-figures-decimal = 3, table-figures-uncertainty = 2]
        S[table-figures-integer = 1, table-figures-decimal = 2, table-figures-uncertainty = 2]
        S[table-sign-mantissa,table-figures-integer = 2, table-figures-decimal = 2, table-figures-uncertainty = 2]
}
    \toprule{} \vspace{-9pt}\\
    {\(m_{\pi}^2 \, / \, \si{\giga\electronvolt^2}\)} & {\(\rsqElectricSnep{} \, / \, \si{\femto\meter\squared}\)} & {\(\rsqMagneticSnep{} \, / \, \mmSnep{}\! \, / \, \si{\femto\meter\squared}\)} & {\(\mmSnep{}\! \, / \, \mu_N\)} \\
    \colrule{} \vspace{-9pt}\\
    0.3884 \pm 0.0113 & -0.032 \pm 0.027 & 0.50 \pm 0.12 & -1.27 \pm 0.11 \\
    0.2654 \pm 0.0081 & -0.049 \pm 0.021 & 0.83 \pm 0.21 & -1.28 \pm 0.10 \\
    0.1525 \pm 0.0043 & 0.016 \pm 0.039 & 0.77 \pm 0.21 & -1.32 \pm 0.12 \\
    0.0784 \pm 0.0025 & -0.039 \pm 0.060 & {\textemdash} & -1.00 \pm 0.29 \\
    0.0285 \pm 0.0012 & -0.022 \pm 0.081 & {\textemdash} & -1.47 \pm 0.62 \\
    \botrule{}
\end{tabular*}
\end{table}

\bigskip

\section{Conclusion\label{sec:excitations:conclusion}}

In this paper we presented the first calculations of the elastic form factors of lattice nucleon
excitations from the first principles of QCD.  We have considered a variety of momentum frames to
access a range of \(Q^2\) values, with some approaching zero.  We have presented the first
implementation of the PEVA technique for matrix elements, which is vital to isolating exited
baryons with non-zero momentum.

Substantial differences between the form factors calculated in a conventional variational analysis
and those calculated with the PEVA approach show the PEVA technique to be critical to
understanding the structure of these excited states.

We find the size of the two lowest-lying negative parity excitations to be similar to the ground
state nucleons.  This is a remarkable result in the context of a simple constituent quark model,
where the repulsive centripetal term of the radial Schr\"odinger equation proportional to $\ell\,
(\ell +1)$ is expected to force the quarks to larger radii for $\ell = 1$ states.  The
lattice QCD results point to a role for meson-nucleon Fock-space components in the $N^*$ states.
As the antiquark in the meson provides the negative parity, all quarks can reside in relative
$s$-waves.  In this way the centripetal barrier is avoided and the negative-parity states can have
a size similar to the ground state nucleon.

The positive-parity excitation observed in this study is very high in energy, approaching \SI{2}{\giga\electronvolt},
and exhibits challenging statistical fluctuations.  The extractions of the form factors herein are
attained through the cancellation of statistical fluctuations enabled by the combination of an
${\cal O}(a)$-improved conserved vector current and an appropriately selected correlator ratio
preserving the lattice Ward identity.  This state has a charge radius\index{charge radius}
approximately \SI{30}{\percent} larger than the ground state, and magnetic moments which match the
ground state.  Both of these observations are in accord with earlier observations that this
positive parity excitation has a wave function consistent with a three-quark $2S$ radial excitation
\cite{Roberts:2013ipa,Roberts:2013oea}.

At the heaviest three pion masses considered, the first observed negative-parity excitations have
magnetic moments consistent with quark-model descriptions for the \(N^*(1535)\).  Similarly, the
second negative-parity excitations have magnetic moments in accord with quark-model descriptions of
the \(N^*(1650)\)\index{N*1650@\(N^*(1650)\)}.  At these quark masses, the
results indicate these states are similar in structure to the ground state nucleon which can
also be modelled as a three-quark state dressed by a meson cloud.

At the lightest two pion masses, we observe a rearrangement in the structure of the second
negative-parity excitation. This is evident in both a significant shift in the magnetic moments of
the excited proton and neutron, and significant curvature in the pion-mass dependence of the
electric form factor. A description of this state as a molecular bound state of \(K \Sigma\)
dressed by \(K \Lambda\), \(\eta N\) and \(\pi N\) is an intriguing possibility, analogous to the
description of the odd-parity \(\Lambda(1405)\) excitation as a molecular bound state of
\(\adjoint{K} N\) dressed by \(\pi \Sigma\)~\cite{Hall:2014uca,Hall:2016kou}. The proximity of the
non-interacting \(K \Sigma\) to the effective energy of the observed lattice state is suggestive.

While the current analysis is able to raise such a possibility, it is not able to affirm it.  As a
first step, non-local momentum-projected two-particle interpolating fields must be introduced to
provide access to the \(K \Sigma\), \(K \Lambda\), \(\eta N\) and \(\pi N\) scattering states.  To
date, only the energy of the $S$-wave \(\pi N\) scattering state has been investigated
\cite{Lang:2012db,Liu:2015ktc}.
This will be a challenging endeavour, owing to the computational cost of estimating the loop
propagators that are necessary to compute non-local momentum-projected meson-baryon contracted
interpolating fields.
Moreover, very high statistics will be required to precisely evaluate the correlation functions for
quark masses near the physical regime.

However, such studies will allow for the lattice determination of the form factors of the
multi-particle dominated scattering-states at light quark masses, and will provide the input
required to make a robust connection between these states and the
infinite-volume resonances of nature \cite{Briceno:2015tza,Baroni:2018iau}.

Looking forward, it is interesting to note that HEFT calculations describe the negative-parity
energy spectrum with reference to one bare basis state \cite{Liu:2015ktc}.  This investigation
indicates the presence of two quark-model like states raising the possibility of introducing a
second bare basis state into the formalism.  It will be interesting to explore such an extension of
HEFT as the two quark-model-like basis states mix through couplings to intermediate meson-baryon
basis states.  Such dynamics may be relevant to a detailed quantitative understanding of the
negative-parity nucleon spectrum.

\appendix*

\begin{acknowledgments}
This research was undertaken with the assistance of resources from the
Phoenix HPC service at the University of Adelaide, the National
Computational Infrastructure (NCI), which is supported by the Australian
Government, and by resources provided by the Pawsey Supercomputing Centre
with funding from the Australian Government and the Government of Western
Australia. These resources were provided through the National Computational
Merit Allocation Scheme and the University of Adelaide partner share. This
research is supported by the Australian Research Council through grants
no.\ DP140103067, DP150103164, LE160100051, LE190100021 and DP190102215.
\end{acknowledgments}

\bibliography{reference}

\end{document}